\documentclass[usenatbib]{mn2e}
\usepackage{rotating}
\usepackage{graphicx}
\usepackage[T1]{fontenc}
\usepackage{times}
\usepackage{amssymb}
\usepackage{amsmath}
\usepackage{textcomp}
\usepackage{url}
\usepackage{rotfloat}
\usepackage[normalem]{ulem}
\newcommand{\lsol}{\textrm{L}_{\odot}}
\newcommand{\msol}{\textrm{M}_{\odot}}
\newcommand{\uitmatches}{348}		
\newcommand{\numOurSources}{27901}	
\newcommand{\uitFUVoffsetavg}{-0.04}	

\newcommand{\uitNUVoffsetavg}{0.73}	

\newcommand{\uitFUVoffsetsigma}{0.60}	
\newcommand{\uitNUVoffsetsigma}{0.38}

\newcommand{\ourFUVapcor}{0.03}		
\newcommand{\ourNUVapcor}{0.06}		
\newcommand{\numOneMagOffFUVuit}{30}	
\newcommand{\numOneMagOffNUVuit}{65}	
\newcommand{\uitFUVfilterdiff}{-0.02}	
\newcommand{\uitNUVfilterdiff}{0.17}	
\newcommand{\raOffset}{0.^{\!\!\prime\prime}17}	  
\newcommand{\decOffset}{-0.^{\!\!\prime\prime}08}

\newcommand{\numoptuvmatches}{24738}	
\newcommand{\numOneMatchoptical}{5032}	
\newcommand{\avgOptMatches}{2.75}			
	
\newcommand{\avgOptOffset}{1.^{\!\!\prime\prime}25}	
\newcommand{\stdevOptOffset}{0.^{\!\!\prime\prime}74}	
\newcommand{\wrMatches}{114}				
\newcommand{\maxOptMatches}{20}				
\newcommand{\nuvMinusVcoloravg}{-0.42}						
\newcommand{\nuvMinusVcolormedian}{-0.45}		
\newcommand{\abmagformula}{m_{\nu}=-2.5\log\left(F_{\lambda}\times\frac{\lambda^{2}}{c}\right)-48.60}

\title{GALEX Catalog of UV Point Sources in M33}
\author[Dale Mudd and K.Z. Stanek ]{\parbox{18cm}{Dale Mudd$^{1}$, K.Z. Stanek$^{1}$}
\\
$^{1}$Dept.\ of Astronomy, The Ohio State University, 140 W.\ 18th   Ave., Columbus, OH 43210\\
E-mail: mudd@astronomy.ohio-state.edu}
\begin{document}

\maketitle

\begin{abstract}
The hottest stars ($>$10,000 K), and by extension typically the most massive ones, are those that will be prevalent in the ultraviolet (UV) portion of the electromagnetic spectrum, and we expect to numerous B, O, and Wolf-Rayet stars to be bright in UV data.  In this paper, we update the previous point source UV catalog of M33, created using the Ultraviolet Imaging Telescope (UIT), using data from the Galaxy Evolution Explorer (GALEX).  We utilize PSF photometry to optimally photometer sources in the crowded regions of the galaxy, and benefit from GALEX's increased sensitivity compared to UIT.  We match our detections with data from the Local Group Galaxies Survey (LGGS, \citealp{Massey06}) to create a catalog with photometry spanning from the far-UV through the optical for a final list of \numoptuvmatches\textrm{} sources.  All of these sources have far-UV (FUV; 1516\AA), near-UV (NUV; 2267\AA), and \emph{V} data, and a significant fraction also have \emph{U}, \emph{B}, \emph{R}, and \emph{I} data as well.  We also present an additional 3000 sources that have no matching optical counterpart.  We compare all of our sources to a catalog of known Wolf-Rayet stars in M33 \citep{Neugent11} and find that we recover \wrMatches\textrm{} of 206 stars with spatially-coincident UV point sources.  Additionally, we highlight and investigate those sources with unique colors as well as a selection of other well-studied sources in M33.  
\end{abstract}

\section{Introduction}
\label{sec: intro}
M33, also known as the Triangulum Galaxy, was "officially" discovered by \citet{Messier81}, although it might have been noted more than a century earlier by \citet{Hodierno1654}. There is a wide range of calculated distances for this Local Group galaxy, ranging from 700-1100 kpc, and we adopt a distance of d = 964 kpc (\citealp{Bonanos06}).  Being so close, it is a well-studied galaxy indeed: as of April 2015, the Astrophysics Data System reveals $>$1,200 refereed articles with ``M33'' in the abstract.  M33 has been studied across many wavelengths (e.g., \citealp{Helfer03, Massey06, Warner73, Pietsch04, Long10, Thompson09, Massey96}) and across time (e.g., \citealp{Hubble26, Hubble53, Freedman91, Macri01, Hartman06}).  Perhaps surprisingly, there have been relatively few studies of M33 in UV wavelengths (e.g., \citealp{Massey96, Thilker05}), and no catalog of point sources based on GALEX images of M33 has been published, a situation which we rectify in this paper.

The Galaxy Evolution Explorer (GALEX) was launched in 2003.  It was designed as a UV all-sky survey, with 5 smaller surveys making up the first portion of the mission.  Specifically, we employ data from the Nearby Galaxy Survey (NGS) in the current work.  The goal of the telescope was to gain a better understanding of galaxy evolution by studying local star formation, star formation histories, extinction, and UV galaxy morphology \citep{Martin03}.  The instrument consisted of a 50 cm telescope connected to two sealed tube detectors and microchannel plates with a peak quantum efficiency of about 10\%.  The dichroic splitter allowed for simultaneous observation in both the near and far UV filters spanning from 1350 to 2800\AA.  The circular field of view on the telescope had a diameter of roughly 1.2\textdegree.  For more on the technical aspects of GALEX, see \citet{Jelinsky03} and \citet{Morrissey07}.

A UV study of M33 is of potentially great importance.  The most massive stars may appear quite faint in an optical survey since most of their emission may be concentrated in shorter wavelength bands.  As such, a mission such as GALEX is ideal for identifying and characterizing these massive stars.  Wolf-Rayet stars are also much more prominent in the UV with their extreme surface temperatures.  In this work, we create a catalog of UV point sources in M33 using the GALEX space telescope.  We match this to a ground-based optical catalog, as well as a list of known Wolf-Rayet stars, in the galaxy and present photometric data spanning seven filters for tens of thousands of sources, which can be used for numerous astrophysical applications.

We begin by discussing the methods used for the construction of our catalog and matching against the previous UV and optical catalogs of M33 in \textsection\ref{sec: data}.  In \textsection\ref{sec: disc}, we discuss the most interesting sources and aspects of our final product and then conclude.

\section{Data and Methods}
\label{sec: data}
\subsection{UV Photometry}
\label{subsec: UVmethod}
We begin with the UV science images\footnote{We retrieved these images from the Barbara A. Mikulski Archive for Space Telescopes (MAST).} of M33 taken with the GALEX space telescope during the NGS on November 11, 2003.  There are several pointings at M33 as part of GALEX's survey.  Tilings around the galaxy exist, but we opt to work with the central pointing of the galaxy since the photometric repeatability level of fainter sources, combining several tilings, could introduce errors of up to 0.4 magnitudes (see \citealp{Morrissey07}).  The two exposures combine for a total exposure time of 3334 seconds.  The simultaneous near-UV (NUV) and far-UV (FUV) images have passbands 1750-2800\AA\textrm{} and 1350-1750\AA\textrm{} and effective wavelengths of 2267\AA\textrm{} and 1516\AA, respectively.  The PSF full-width half-maximum (FWHM) for the NUV and FUV detectors is $4.^{\!\!\prime\prime}2$ and $5.^{\!\!\prime\prime}3$, respectively, sampled with $1.^{\!\!\prime\prime}5$/pixel.

\begin{figure}
  \centerline{
    \includegraphics[width=8.0cm]{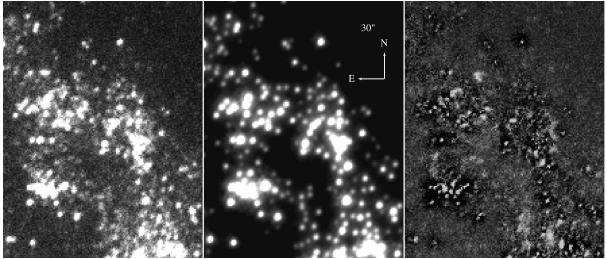}
}  
  \caption{From left to right, the three panels are a fraction of a spiral arm from the original data, the sources detected, and the remaining flux after subtracting these sources using the derived PSF.  The spatial scale is shown in the middle panel, where each line on the compass corresponds to $30^{\prime\prime}$, as shown in the figure.}
  \label{fig: subtraction}
\end{figure}

The $\textrm{D}_{25}$, the diameter at which the B band surface brightness drops to 25 mag/$\textrm{arcsec}^{2}$, for M33 is $1.2$\textdegree\textrm{} \citep{deVaucouleurs91}, which corresponds well to the GALEX field of view.  We began construction of our point source catalog with the ``int'' images from the GALEX pipeline, in units of counts per second corrected for effective exposure times, by performing PSF photometry with a combination of the DAOPHOT and ALLSTAR programs \citep{Stetson87} on the NUV and FUV images separately.  In our reduction, sources are required to have at least a 5$\sigma$ detection with less than 0.3 magnitudes in their uncertainty, and we begin by fitting sources with a purely analytic PSF for our baseline processing, a step that is strongly suggested for crowded sources as is the case with M33.  A subset of the brightest and relatively isolated sources found under these conditions are then re-run through the reduction after subtracting out their nearest neighbors to iteratively improve the analytical PSF model empirically.  In general, the empirical PSFs were slightly more elliptical than the analytic version.  This refined PSF became the basis for our final source extraction.  As an illustration, we show a comparison between the original image, the recovered stars, and the difference between the two in Figure \ref{fig: subtraction}.  From this figure, it is evident that many sources are successfully detected, but the limitation caused by the crowding in dense regions of the galaxy is readily apparent.

We then derived the necessary aperture correction on the brightest few hundred sources by using successively larger apertures (after subtracting out all other detected sources) and measuring the magnitude offsets between these and those measured with the final PSF model.  For our aperture corrections, we found values of \ourFUVapcor\textrm{} magnitudes for the FUV and \ourNUVapcor\textrm{} magnitudes for the NUV.  We next converted our instrumental magnitudes back to counts per second, which have been calibrated into both fluxes and AB magnitudes \citep{Hayes75} by the GALEX Team \citep{Morrissey07}.  The AB magnitude system, in wavelength, is defined as
\begin{equation}
 \abmagformula,
 \label{eq: ABmag}
\end{equation}
where flux $F_{\lambda}$ is given in $\textrm{erg s}^{-1}\textrm{cm}^{-2}\textrm{\AA}^{-1}$.   

After applying the respective aperture corrections to the NUV and FUV catalogs and converting all sources to the AB system, we sought to match the FUV and NUV source catalogs.  To do so while minimizing false matches, we began by matching using a 3 pixel maximum radius, or approximately $4.^{\!\!\prime\prime}5$.  As there were fewer sources in the FUV, we then kept only those sources with only one NUV match to a given FUV source within this radius.  With this stringent cut of relatively large matching radius combined with only a single match, this list would contain a high concentration of correct source pairings.  From these matches, we found the average offset between the NUV and FUV magnitudes (i.e. colors) and positions, along with the dispersion around these two values.  These were then used to do a second round of matching the NUV to the FUV source list, keeping sources with more than a single counterpart in the NUV this time, based on minimizing the distance between objects in position-magnitude space with a maximum allowed physical separation of 2 pixels, or $\sim3^{\prime\prime}$.  This resulted in a catalog of \numOurSources\textrm{} distinct GALEX FUV sources with NUV matches.  Matching this way rather than through distance alone changed 209 total matches.  Compared to the distance-only matches, our matches are of comparable separation, different by about $0.^{\!\!\prime\prime}04$.  However, they tend to be about 1 magnitude reduced in UV color, which removed many strong color outliers from the catalog.

\begin{figure}
  \centerline{
    \includegraphics[width=8.0cm]{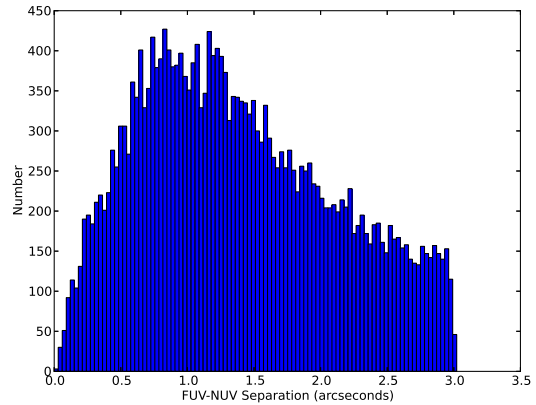}
}  
  \caption{Histogram of the absolute separation, in arcseconds, between a FUV source and its corresponding NUV match.}
  \label{fig: fnuv_sep_hist}
\end{figure}

\begin{figure}
  \centerline{
    \includegraphics[width=8.0cm]{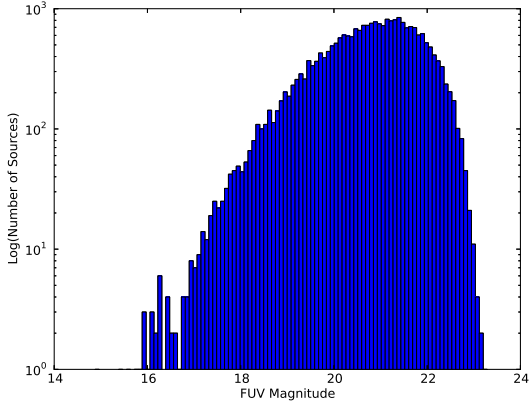}
}  
  \caption{The FUV luminosity function of our final matched set of UV sources.}
  \label{fig: fuv_lumin_func}
\end{figure}

\begin{figure}
  \centerline{
    \includegraphics[width=8.0cm]{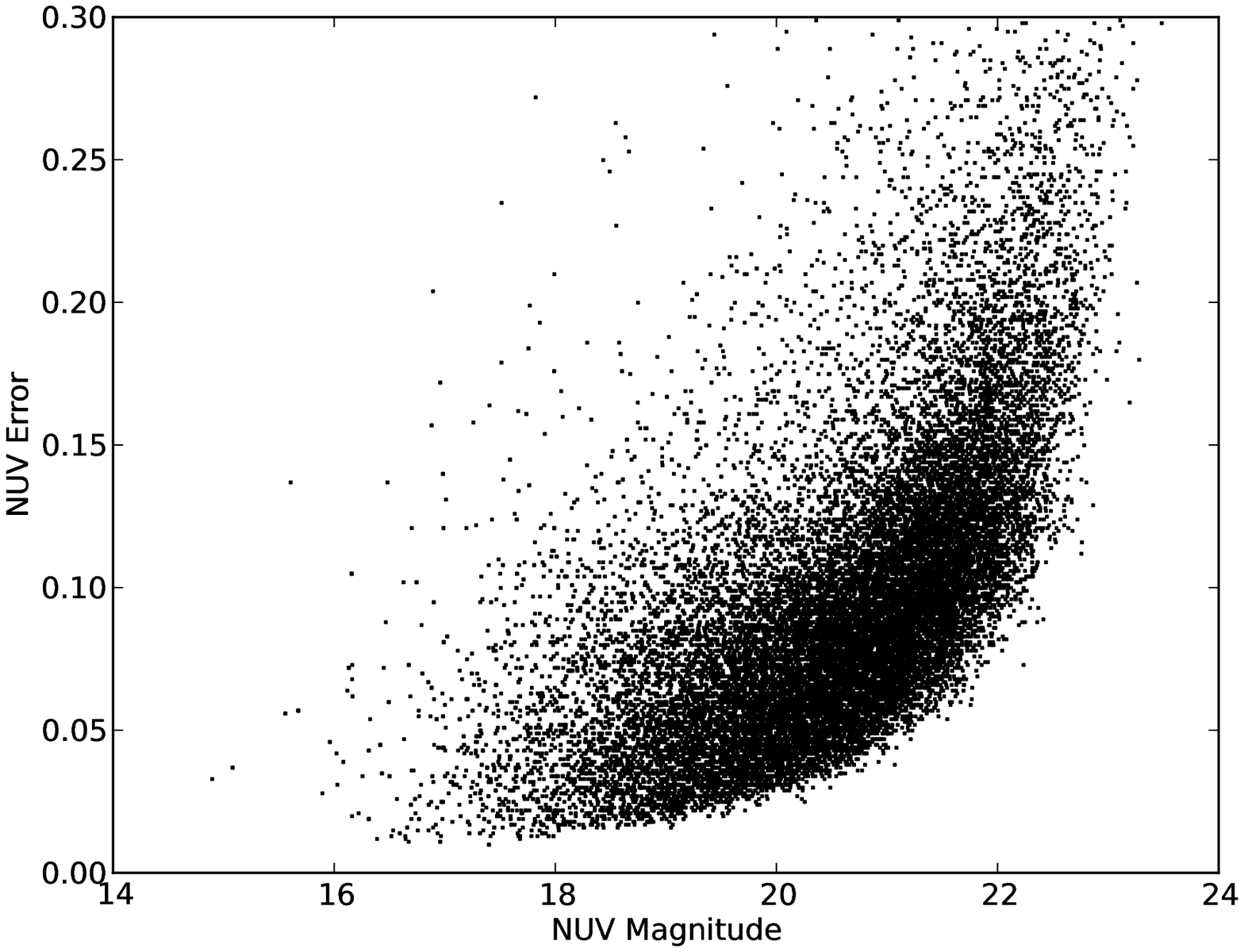}
}
  \centerline{
    \includegraphics[width=8.0cm]{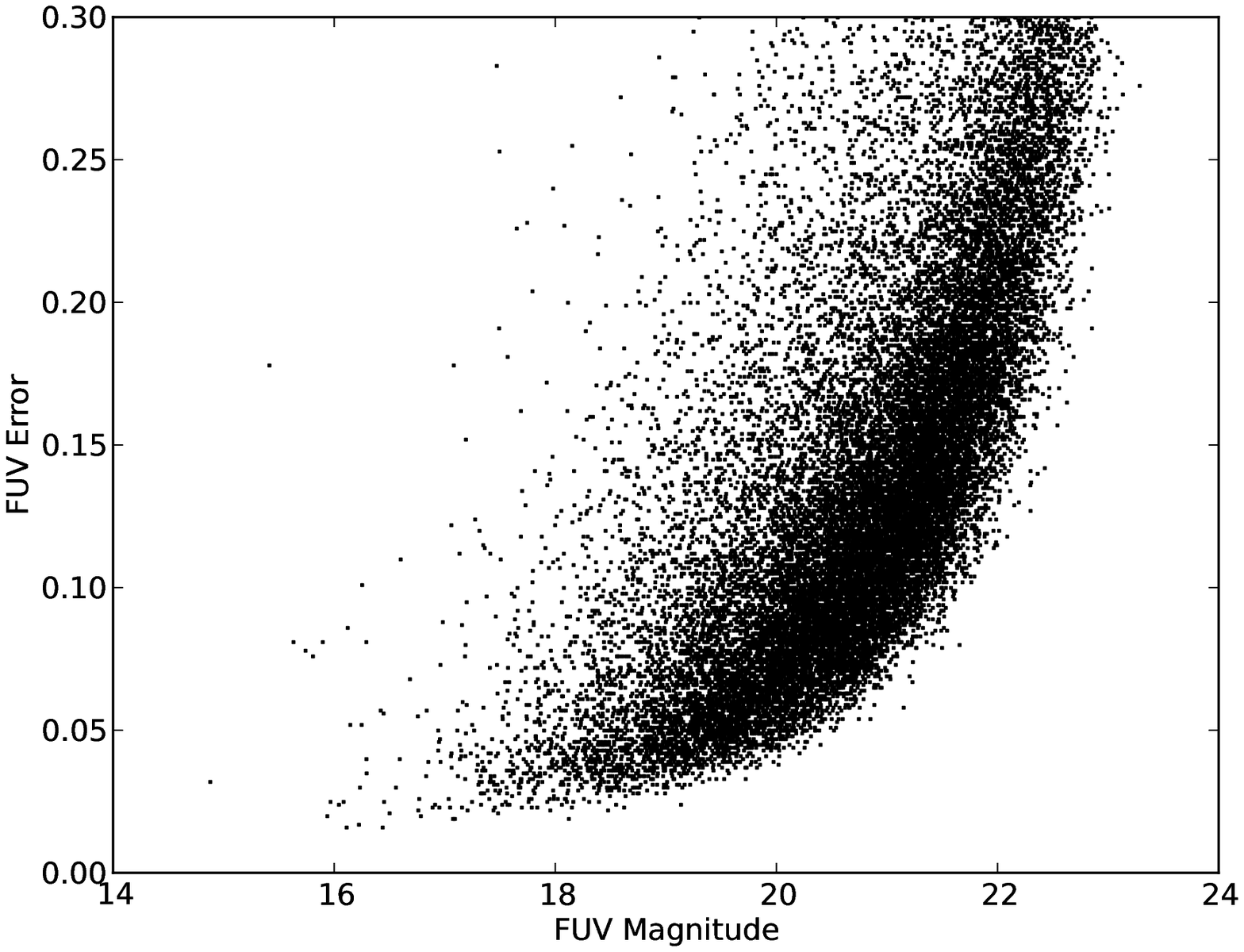}
}  
  \caption{Error as a function of magnitude in the NUV (\emph{top}) and FUV (\emph{bottom}).  As one might expect, the errors tend to grow for both filters for fainter sources.}
  \label{fig: uv_v_err}
\end{figure}

\begin{figure}
  \centerline{
    \includegraphics[width=8.0cm]{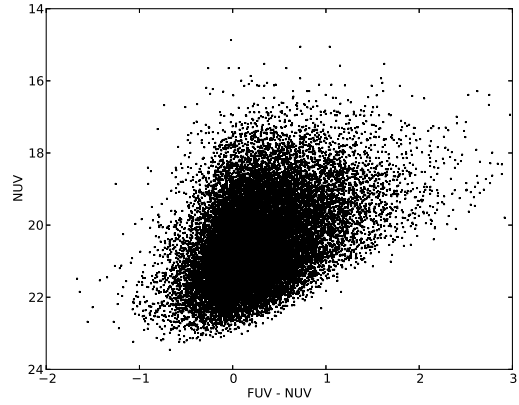}
}  
  \caption{UV color-magnitude diagram of the final sources in our catalog.}
  \label{fig: uv_cmd_2}
\end{figure}

We then investigated the gross properties of our matches.  We compare the source locations in the NUV to the FUV in Figure \ref{fig: fnuv_sep_hist}.  This gives us a sense of the spatial separation of our sources as well as their positional uncertainties, which are somewhat large.  Figure \ref{fig: fuv_lumin_func} shows the FUV luminosity function of our sources, which steadily rises until a turnover around 21 mag.  Errors as a function of magnitude in both NUV and FUV are shown in Figure \ref{fig: uv_v_err} and a UV color-magnitude diagram is presented in Figure \ref{fig: uv_cmd_2}.

\begin{figure}
  \centerline{
    \includegraphics[width=8.0cm]{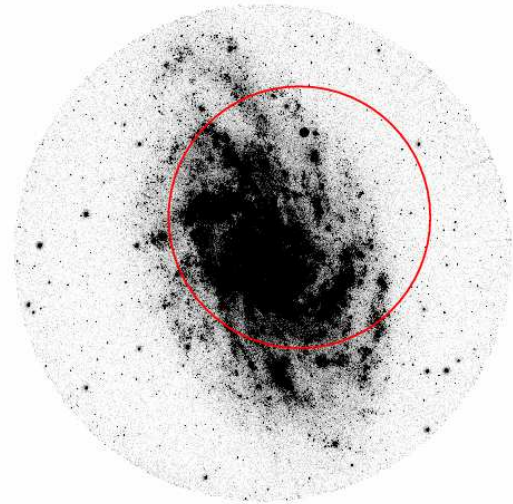}
}  
  \caption{GALEX near-UV image of M33.  The inner circular region (radius of $18^{\prime}$) is the field of view of the Ultraviolet Imaging Telescope (UIT), used to create the previous UV point source catalog for M33 \citep{Massey96}.}
  \label{fig: fovComp}
\end{figure}

Next, we compared our catalog to the existing UV catalog of M33 sources compiled in \citet{Massey96}.  This catalog was made using the Ultraviolet Imaging Telescope (UIT), an instrument aboard the Astro-1 Mission \citep{Stecher92}.  UIT used photographic plates with the B1 and A1 filters roughly corresponding to the FUV and NUV filters of GALEX, having central wavelengths of $\sim$1500\AA\textrm{} and 2400\AA, respectively.  It should be noted, however, that the A1 filter is significantly broader than the NUV filter on GALEX, reaching several hundred angstroms to the red end of its GALEX counterpart.  The field of view of UIT is also circular but has a smaller radius of $18^{\prime}$, which can be seen within the GALEX field of view in Figure \ref{fig: fovComp}.  The FWHM of UIT is comparable to that of GALEX, at $4^{\prime\prime}$ and $5.^{\!\!\prime\prime}2$ in the NUV and FUV filters, respectively.  GALEX, however, reaps the reward of technological advancement over time in its implementation of more efficient microchannel plates instead of UIT's photographic ones.  For more information on UIT and its detector's properties, see \citet{Stecher92} and \citet{Landsman92}.  

\begin{figure*}
  \centerline{
    \includegraphics[width=14.0cm]{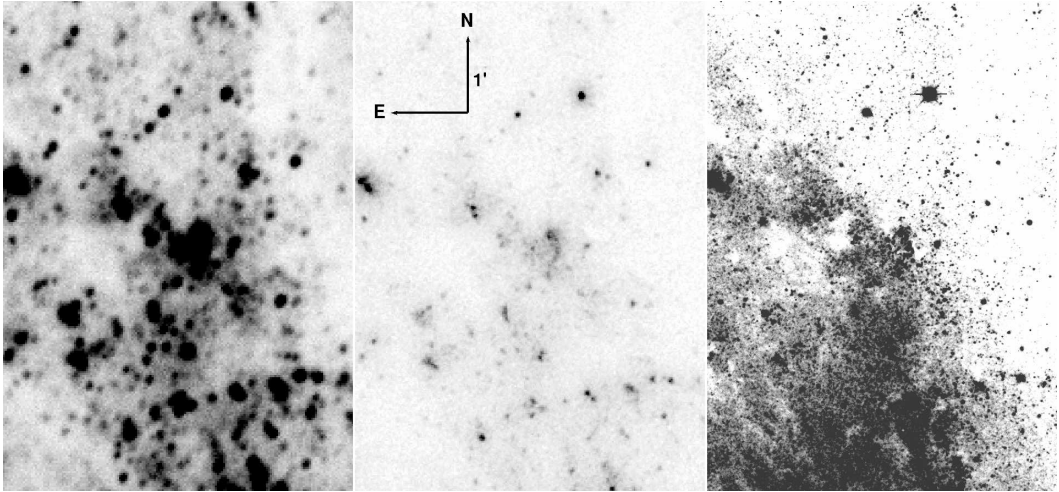}
}  
  \caption{Comparison of a region in M33 with GALEX (\emph{Left}), UIT (\emph{Middle}), and V-band from the Local Group Galaxies Survey with the Mayall 4m telescope (\emph{Right}, \citealp{Massey06}).  The two UV instruments have similar full-width half-maxima ($\sim4^{\prime\prime}$), while GALEX has much greater sensitivity.  The sides of the compass in the middle figure correspond to a size on the sky of $1^{\prime}$.  The optical image highlights the problem of crowding that is prevalent throughout the UV data.}
  \label{fig: depthComp}
\end{figure*}

Similar to the analysis route we adopted, \citet{Massey96} only keep sources for which there are both NUV and FUV detections, and they supplement these with \emph{U}, \emph{B}, and \emph{V} ground-based data as well.  Their final catalog has 356 sources (note that the catalog naming scheme goes as high as 374 as certain numbers are skipped), which they judge to be complete to an FUV AB magnitude of $\sim$18.5, corresponding to a specific flux of $F_{1500\textrm{\AA}} = 2.5\times10^{-15}\textrm{ erg cm}^{-2}\textrm{s}^{-1}\textrm{\AA}^{-1}$, and its faintest source has an FUV magnitude of 19.7, corresponding to a specific flux of $F_{1500\textrm{\AA}} = 6.4\times10^{-16}\textrm{ erg cm}^{-2}\textrm{s}^{-1}\textrm{\AA}^{-1}$.  The GALEX observations are much more sensitive, and over 20,000 of our sources are fainter than this, a difference that is highlighted in Figure \ref{fig: depthComp}.  The faintest source in our final catalog has an FUV magnitude of 23.3, corresponding to a specific flux of $F_{1500\textrm{\AA}} = 2.3\times10^{-17}\textrm{ erg cm}^{-2}\textrm{s}^{-1}\textrm{\AA}^{-1}$.  

\begin{figure}
 \centerline{
 \includegraphics[height=6.0cm]{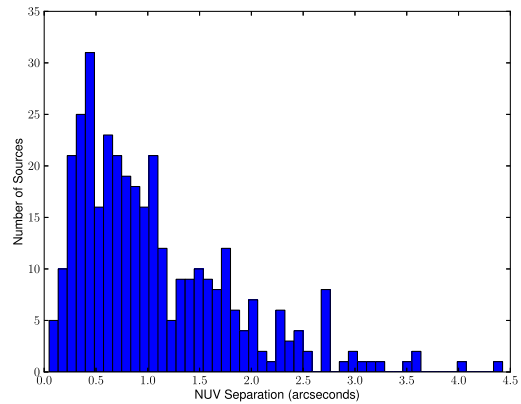}
} 
 \caption{Distribution of matches between UIT \citep{Massey96} and our GALEX catalog as a function of source separation in the near-UV (NUV) bands.}
 \label{fig: separations}
\end{figure}

Looking at the astrometry solution for GALEX, we found there were some slight distortions in the two images, especially near the edges of the galaxy.  We use the astrometry.net world coordinate system (WCS) solution for our field \citep{Lang12} to correct this\footnote{These images with the new WCS solutions can be found at \url{http://www.astronomy.ohio-state.edu/~mudd/gallery.html}}.  As the FUV image image does not have enough sources outside the galaxy to find a solution, we apply the solution for the NUV image to the FUV one as well.  Similar to the procedure followed for matching our NUV and FUV sources internally, we matched to the UIT catalog, this time with a radius of $3.^{\!\!\prime\prime}2$ to complement our previous maximum radius, cut to those sources for which there was only a single UIT star within this radius, and then found the average and standard deviation for the magnitude and position offsets between the two catalogs.  After this, we rematched based on magnitude and position once again.  The distribution of matches by separation is shown in Figure \ref{fig: separations}.  From this, we decided to stick with a maximum radial separation of $3.^{\!\!\prime\prime}2$ as before.  The remaining sources making this cut had a right ascension and declination offsets (UIT - GALEX) of $\raOffset$ and $\decOffset$, respectively.  Both of these are quite small compared to the size of the PSF. 

\begin{figure*}
 \centerline{
   \includegraphics[width=8.0cm]{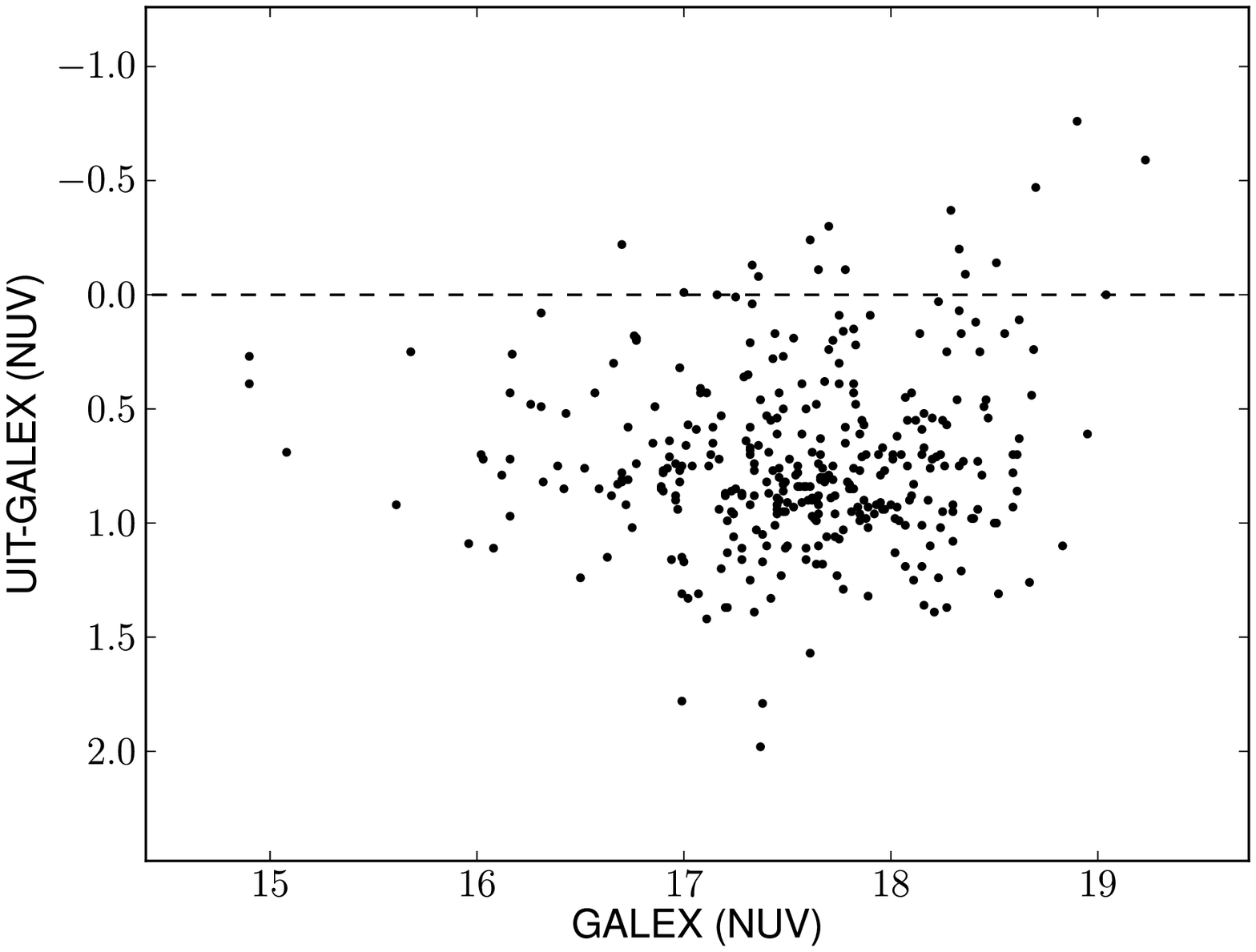}
   \includegraphics[width=8.0cm]{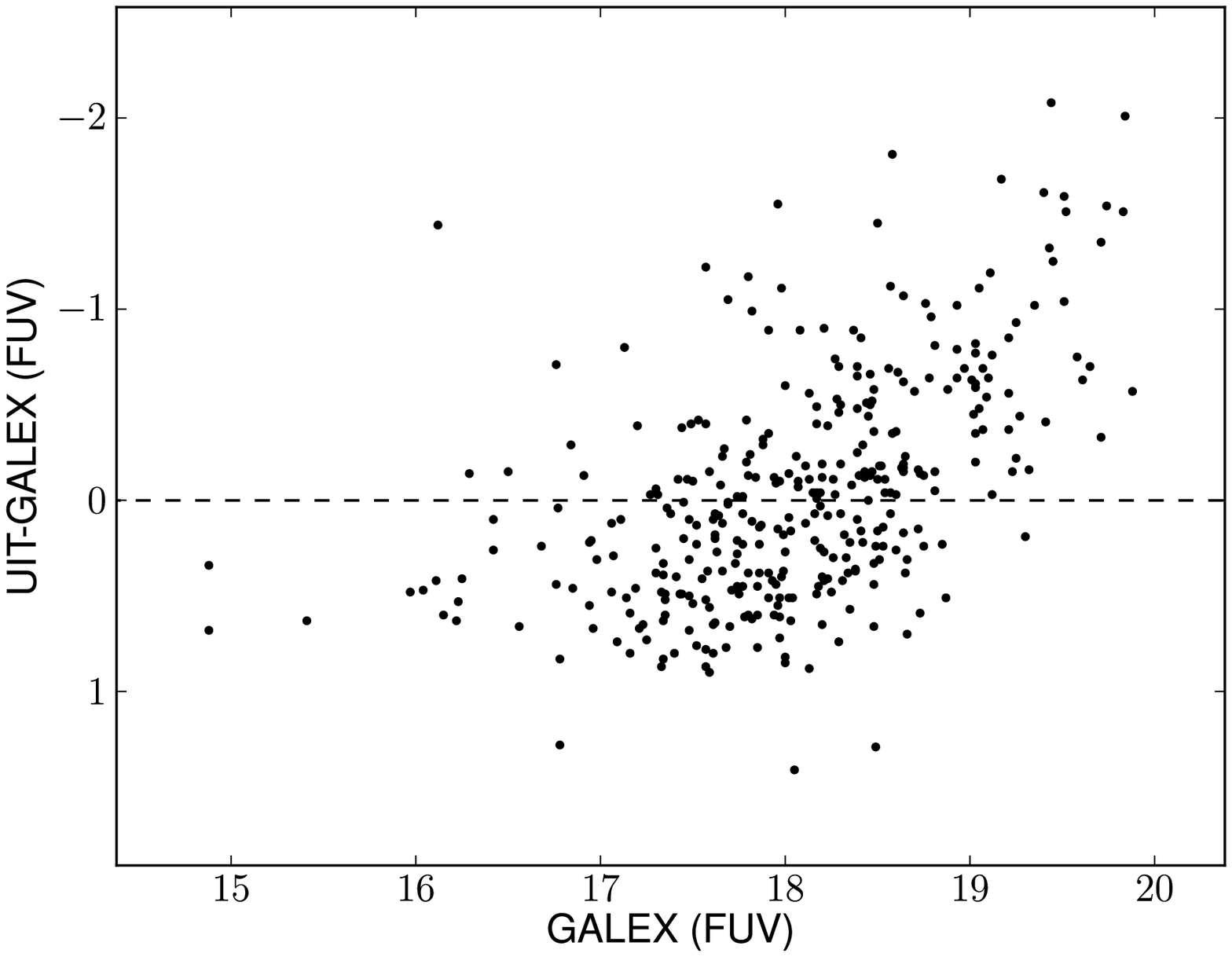}
 }
 \caption{AB magnitude difference between UIT \citep{Massey96} and our GALEX catalog as a function of GALEX NUV (\emph{left}) and FUV (\emph{right}) magnitude.}
 \label{fig: mag_differences}
\end{figure*}

This final catalog had \uitmatches\textrm{} sources in common with the original UIT catalog of M33, which contained 356 sources.  The average difference between NUV magnitudes (UIT - GALEX) is \uitNUVoffsetavg\textrm{} mag with a spread of \uitNUVoffsetsigma.  For the FUV filters, the same calculation gives an offset of \uitFUVoffsetavg\textrm{} mag and a dispersion of \uitFUVoffsetsigma.  These offsets as a function of magnitude in both NUV and FUV are shown in Figure \ref{fig: mag_differences}.  While the NUV has a large offset, it does not appear to have any structure in it, whereas the FUV shows a trend of the GALEX photometry of faint sources being systematically fainter than in the original catalog.  Looking at the eight sources that we did not recover, seven of them were in crowded regions separated into sources by our PSF photometry differently than UIT such that the UIT object had several potential GALEX matches surrounding it, but none were within the matching radius.  The final source, [MBH96] 167, is in a more diffuse, fainter region and it is surprising that it does not appear in our catalog given that GALEX has a higher sensitivity than UIT.  We investigate our raw photometry and the images themselves and found this object appears in both of the images and in the initial far-UV photometry but was not selected as a near-UV source and hence did not make it into our catalog.  

We inspected by eye many of the sources with the largest discrepancies in magnitude ($\Delta$ mag\textgreater1) in either band between the UIT and our catalog.  The largest offsets in the FUV were typically sources that were in crowded regions that were split into several sources in GALEX and tended to be brighter in UIT, as one might expect.  The largest offsets in the NUV, however, were all brighter in our catalog.  Most of these were in more diffuse or isolated regions compared to the more clustered FUV-discrepant sources.  There were a total of \numOneMagOffNUVuit, \numOneMagOffFUVuit\textrm{} sources differing by more than a magnitude in the NUV and FUV, respectively.  

To test for any systematic separation between our magnitudes and those from \citet{Massey96} due to differences in the filter shapes of UIT and GALEX, we applied the four filter curves (UIT A1, B1 and GALEX FUV, NUV) to various temperature black bodies to estimate what the magnitude separation would be between UIT and GALEX.  The FUV filters for the two instruments are quite similar, and, as such, the magnitude difference for a 40,000 K black body (UIT - GALEX), a temperature equivalent to a warm O star, was \uitFUVfilterdiff, which is quite close to the average offset we reported above.  For the NUV, however, the 40,000 K black body has a magnitude difference (again, UIT - GALEX) of \uitNUVfilterdiff, which does not fully account for the average NUV offset we find.  The exact cause of this offset remains unclear.  Given the extensive testing of the GALEX detectors (see again \citealp{Morrissey07}), however, we are more inclined to trust the new UV magnitudes of the sources presented here.

The UIT catalog, with both the original and newly-derived GALEX parameter values (when available), is presented in Table 1.

\pdfminorversion=5
\begin{sidewaystable}
\centering
{GALEX Photometry of [MBH96] Sources} \\
\begin{tabular}{lrrrrrrrrrrrr}
  \hline
  \hline
  {UIT} & {$\alpha_{\textrm{U}}$} & {$\delta_{\textrm{U}}$} & {$\alpha_{\textrm{G}}$} & {$\delta_{\textrm{G}}$} & {$\textrm{FUV}_{\textrm{U}}$} & {$\textrm{NUV}_{\textrm{U}}$} & {$\textrm{FUV}_{\textrm{G}}$} & {$\textrm{NUV}_{\textrm{G}}$} & {$\textrm{FUV}_{\textrm{U-G}}$} & {$\textrm{NUV}_{\textrm{U-G}}$} & {$\textrm{Sep(}^{\prime\prime}$)} & {\textrm{Matches}}\\
  \hline
2 & 23.13300 & 30.58275 & 23.13335 & 30.58262 & 18.19 & 18.31 & 18.03$\pm$0.06 & 17.62$\pm$0.03 & 0.09 & 0.69 & 1.18 & 3\\ 
3 & 23.15708 & 30.66825 & 23.15677 & 30.66829 & 17.44 & 17.35 & 17.59$\pm$0.08 & 16.86$\pm$0.02 & 0.01 & 0.49 & 0.97 & 2\\ 
4 & 23.15854 & 30.66797 & 23.15813 & 30.66810 & 17.70 & 18.31 & 17.69$\pm$0.06 & 17.48$\pm$0.03 & -0.94 & 0.83 & 1.35 & 1\\ 
5 & 23.17871 & 30.64644 & 23.17855 & 30.64642 & 18.15 & 17.84 & 18.26$\pm$0.15 & 16.96$\pm$0.01 & 1.37 & 0.88 & 0.5 & 2\\ 
6 & 23.18583 & 30.58325 & 23.18597 & 30.58333 & 17.67 & 17.93 & 17.80$\pm$0.09 & 17.40$\pm$0.08 & -0.35 & 0.53 & 0.52 & 2\\ 
\hline
\hline
\end{tabular}%
\\\textbf{Table 1: }UIT sources that were recovered using GALEX data.  Here, 'Matches' refers to the number of GALEX objects we find that are within $5^{\prime\prime}$ and compatible with the given UIT source.  The subscript 'U' stands for UIT, whereas the subscript 'G' stands for GALEX.  The first five sources are given to show the formatting.  The full version of this table is available in the online version. \hfill%
\label{tab: UITmatches}
\end{sidewaystable}
	

\subsection{UV-Optical Catalog}
\label{subsec: opticalmethod}
Having completed our update of the UIT catalog of known M33 UV sources with GALEX photometry, we next set out to expand on this with our additional $\sim$25,000 sources.  To this end, we combine our UV sources with optical counterparts found in a similar-depth study of M33 in U, B, V, R, and I from the Local Group Galaxies Survey (LGGS; \citealp{Massey06}).  

\begin{figure}
 \centerline{
 \includegraphics[width=10.0cm]{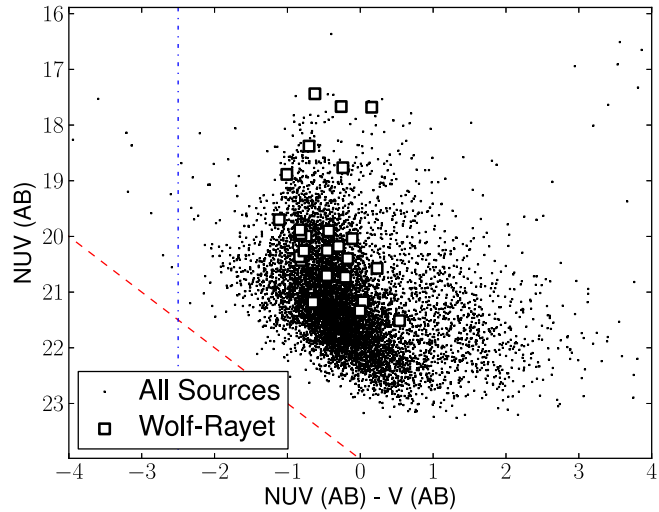}
}
 \caption{Color-magnitude diagram (NUV-V) for GALEX matches to those sources found in the LGGS \citep{Massey06} where only a single optical counterpart was found within $3.^{\!\!\prime\prime}2$ (\numOneMatchoptical\textrm{} sources).  The red dashed line is the approximate magnitude cutoff of the optical study ($\textrm{V}=24$), whereas the blue vertical dot-dashed line is the bluest a blackbody can be if both the NUV and V fluxes lie on the Rayleigh-Jeans tail of a blackbody.  The squares represent the 25 Wolf-Rayet stars, taken from \citet{Neugent11}, that have only a single counterpart.}
 \label{fig: 1matchCMD}
\end{figure}

First, we matched our GALEX catalog to the LGGS catalog with our previous matching radius of $3.^{\!\!\prime\prime}2$.  As before, to get an idea of what unique physical matches look like, we investigated those sources that only had a single optical counterpart within the matching radius.  This informed us of the true sky separations and colors we might expect from all real matches, since these sources are relatively isolated and there will be little confusion in any matching for them.  These single matches are shown in Figure \ref{fig: 1matchCMD}.  Of special note is the blue vertical line in the figure.  This represents approximately the bluest color a blackbody can have (NUV - V = -2.5), assuming both the NUV and V data lie on the Rayleigh-Jeans tail on the red side of the Wien's Law peak.  Stars can, of course, get around this via emission and absorption, but it is illustrative to know what fraction of our sources are ``unphysically'' blue.  In this subsample, we find that only 0.1\% of our pairs are bluer than this line.  The average color (NUV-V) of all our unique-matched sources is -0.09, but the dispersion is not surprisingly high, given the wealth of different stars we are probing, at 0.82.

\begin{figure}
 \centerline{
 \includegraphics[width=10.0cm]{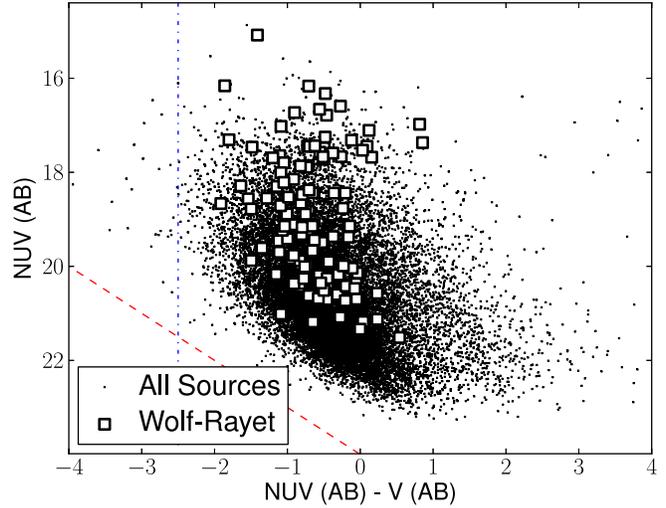}
}
 \caption{Final color-magnitude diagram (NUV-V) for GALEX matches to those sources found in \citet{Massey06} (\numoptuvmatches\textrm{} sources, 114 Wolf-Rayet stars).  The red and blue lines as well as the large squares are the same as those in Figure \ref{fig: 1matchCMD}.  See \ref{subsec: opticalmethod} for a description of the criteria applied to find these matches.}
 \label{fig: allmatchCMD}
\end{figure}

Enlightened by our clean, 1:1 sample, we then iteratively refined our matching algorithm, i.e. combinations of colors and positional offsets, to determine our final matches.  We expected that a successfully matched paired catalog would have similar properties to that shown in Figure \ref{fig: 1matchCMD}.  The final catalog is matched based around a standard deviation of position of $1.^{\!\!\prime\prime}2$, with a maximum allowed separation of $3.^{\!\!\prime\prime}2$, and standard deviation of NUV - V color of 0.8.  The full matched CMD is shown in Figure \ref{fig: allmatchCMD}.  We note that it is bluer on average than the clean sample at \nuvMinusVcoloravg\textrm{} mag with a median of \nuvMinusVcolormedian\textrm{} mag, has slightly lower dispersion, and the same fraction of ``too blue'' stars.

Investigating some of the brightest of these ``too blue'' sources, a pattern emerges.  The brightest of these are all surrounding bright foreground objects and are thus likely being affected by image processing artifacts and their proximity to these extremely bright stars.  The rest appear to cluster in a single dense region of the galaxy, the HII region NGC 604.

Our final catalog, with data spanning 1516-7980\AA\textrm{} for most stars, is provided in Table 2 and has a total of \numoptuvmatches$\textrm{}$ sources.  Note that the catalog has been constructed in such a way that each detected FUV source appears only once, but that each optical source has the freedom to match to more than one FUV source.  When this happens, we report all the separate UV sources that have each given optical star as its best match.  The average number of optical matches to a given UV source is \avgOptMatches$\textrm{}$ with a maximum of \maxOptMatches.  The average positional offset between UV and V coordinates is $\avgOptOffset$ with a standard deviation of $\stdevOptOffset$.  These are both smaller than the FWHM of the GALEX detector.  We also calculate approximate luminosities in each band by using the effective wavelengths of each filter, the GALEX calibrations for the near- and far-ultraviolet flux zeropoints and optical zeropoints taken from \citet{Bessell98}, and the distance to M33 adopted from \citet{Bonanos06}.  Sources without a matching optical counterpart are provided in Table 3.

\pdfminorversion=5
\begin{sidewaystable*}\tiny
{Full Catalog}\\
\begin{tabular}{lrrrrrrrrllllllllllrr}
\hline
\hline
{Name} & {$\alpha_{\textrm{UV}}$} & {$\delta_{\textrm{UV}}$} & {$\alpha_{\textrm{Opt}}$} & {$\delta_{\textrm{Opt}}$} & {$\textrm{FUV}$} & {$\textrm{err}_{\textrm{F}}$} & {$\textrm{NUV}$} & {$\textrm{err}_{\textrm{F}}$} & {$\textrm{U}$} & {$\textrm{err}_{\textrm{U}}$} & {$\textrm{B}$} & {$\textrm{err}_{\textrm{B}}$} & {$\textrm{V}$} & {$\textrm{err}_{\textrm{V}}$} & {$\textrm{R}$} & {$\textrm{err}_{\textrm{R}}$} & {$\textrm{I}$} & {$\textrm{err}_{\textrm{I}}$} & {Sep('')} &  {Matches} \\
\hline

J013239.46+301952.5 &  23.16439 &  30.33169 &  23.16442 &  30.33125 &  21.869 &  0.127 &  21.862 &  0.115 &    22.229 &     0.021 &    22.227 &     0.032 &    22.608 &  0.051 &    22.677 &       0.1 &    99.999 &    99.999 &  1.587 &      1 \\
J013239.45+302241.6 &  23.16430 &  30.37832 &  23.16437 &  30.37822 &  20.082 &  0.056 &  20.157 &  0.041 &    20.425 &     0.004 &    20.616 &     0.009 &    20.891 &  0.013 &    21.049 &     0.012 &    21.284 &     0.004 &  0.421 &      2 \\
J013239.46+303955.1 &  23.16413 &  30.66541 &  23.16442 &  30.66531 &  20.340 &  0.087 &  19.884 &  0.037 &    20.063 &     0.004 &    20.225 &     0.006 &    20.503 &  0.008 &     20.79 &      0.01 &    21.121 &       0.0 &  0.967 &      1 \\
J013354.22+304410.5 &  23.47561 &  30.73663 &  23.47592 &  30.73625 &  20.536 &  0.179 &  20.303 &  0.198 &    19.766 &     0.003 &    19.968 &     0.006 &    20.257 &  0.009 &    20.492 &     0.011 &    20.712 &       0.0 &  1.671 &      2 \\
J013239.45+302214.4 &  23.16448 &  30.37081 &  23.16437 &  30.37067 &  20.438 &  0.061 &  20.514 &  0.061 &    20.563 &     0.005 &    20.623 &     0.012 &    20.858 &  0.012 &    21.114 &     0.016 &    21.456 &       0.0 &  0.609 &      1 \\
J013349.96+304323.4 &  23.45809 &  30.72294 &  23.45817 &  30.72317 &  20.662 &  0.183 &  19.763 &  0.058 &    21.578 &      0.02 &    21.149 &     0.025 &    21.271 &   0.03 &    20.975 &     0.023 &     20.17 &       0.0 &  0.864 &      3 \\
J013349.96+305256.2 &  23.45845 &  30.88262 &  23.45817 &  30.88228 &  18.714 &  0.093 &  18.665 &  0.037 &    19.546 &     0.014 &    19.684 &      0.01 &    19.904 &  0.014 &    19.947 &     0.017 &    20.059 &     0.001 &  1.499 &      4 \\
J013349.96+305256.2 &  23.45845 &  30.88262 &  23.45817 &  30.88228 &  20.451 &  0.104 &  18.665 &  0.037 &    19.546 &     0.014 &    19.684 &      0.01 &    19.904 &  0.014 &    19.947 &     0.017 &    20.059 &     0.001 &  1.499 &      4 \\
J013349.96+303427.8 &  23.45801 &  30.57427 &  23.45817 &  30.57439 &  20.638 &  0.083 &  20.541 &  0.083 &      20.6 &     0.004 &    20.821 &     0.011 &    21.138 &  0.014 &     21.45 &     0.019 &    21.743 &       0.0 &  0.658 &      2 \\
J013321.23+304042.0 &  23.33814 &  30.67849 &  23.33846 &  30.67833 &  20.676 &  0.084 &  20.413 &  0.058 &    20.678 &     0.006 &    20.563 &     0.009 &    20.921 &  0.013 &    21.236 &     0.015 &    21.665 &       0.0 &  1.146 &      2 \\
J013358.39+304823.7 &  23.49306 &  30.80683 &  23.49329 &  30.80658 &  20.493 &  0.122 &  20.091 &  0.044 &    20.702 &     0.006 &    20.674 &     0.011 &    20.905 &  0.014 &      21.1 &     0.018 &    21.386 &       0.0 &  1.147 &      2 \\
 \hline
 \hline
\end{tabular}
\\
\textbf{Table 2: }Our UV sources that had optical counterparts from the LGGS.  Note that all magnitudes are in the AB system.  Here, 'Matches' refers to the number of optical objects we find that are within $3.^{\!\!\prime\prime}2$ of the GALEX source.  Note that 'Sep' refers to the absolute separation between the \\ object's position in the UV and optical catalogs.  The full version of this table is available in the online version.\hfill
\label{tab: Optmatches}
\end{sidewaystable*}

\pdfminorversion=5
\begin{center}\small
\begin{tabular}{lrrrrr}
\hline
\hline
{ID} & $\alpha_{\textrm{UV}}$ & $\delta_{\textrm{UV}}$   &  {$\textrm{FUV}$} & {$\textrm{NUV}$} & $\textrm{N}_{\textrm{mat}}$ \\
\hline
6     &  23.62497 &  30.10737 &  15.713 $\pm$0.022 &  16.135  $\pm$  0.02 &    2 \\
363   &  22.94542 &  30.36758 &  17.978 $\pm$0.035 &  16.992  $\pm$ 0.132 &    2 \\
396   &  23.67746 &  30.42422 &  18.039 $\pm$ 0.062 &  14.314 $\pm$ 0.026 &    1 \\
525   &  24.09594 &  30.67127 &  18.213 $\pm$ 0.043 &  15.853 $\pm$ 0.022 &    2 \\
711   &  23.96306 &  30.32453 &   18.38 $\pm$ 0.045 &  15.622 $\pm$ 0.015 &    1 \\
759   &  23.00021 &  30.36543 &  18.413 $\pm$ 0.041 &  15.022 $\pm$ 0.036 &    1 \\
\hline
\hline
\end{tabular}
\\
\textbf{Table 3:}  Those UV sources that did not have an optical counterpart.  Here, $\textrm{N}_{\textrm{mat}}$ refers to the number of NUV sources within $3.^{\!\!\prime\prime}2$ of the FUV source.  The full version of this table is provided in the online version.\hfill
\end{center}

\section{Discussion}
\label{sec: disc}
\begin{figure*}
  \centerline{
    \includegraphics[width = 7.0cm]{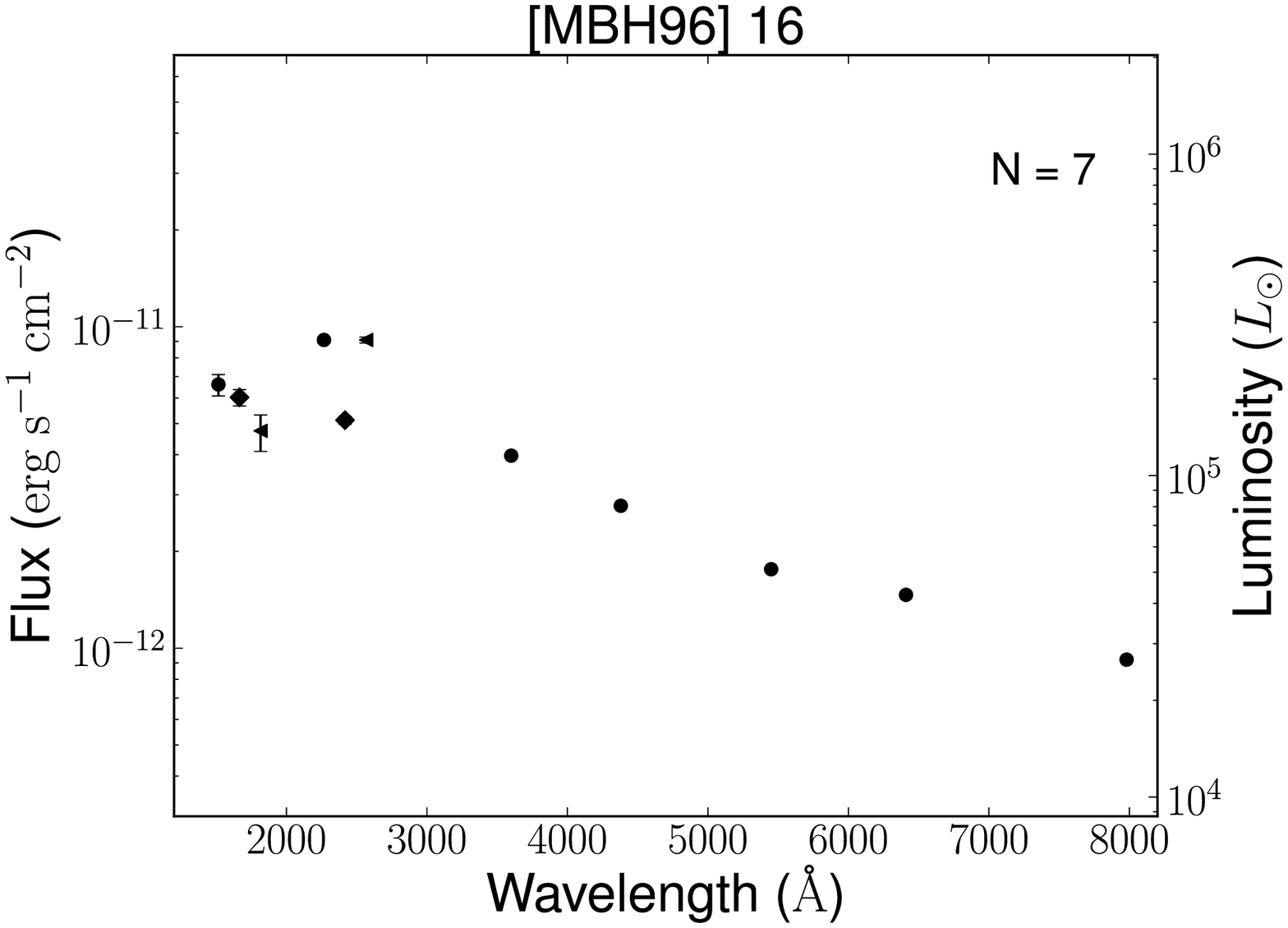}
    \includegraphics[width = 7.0cm]{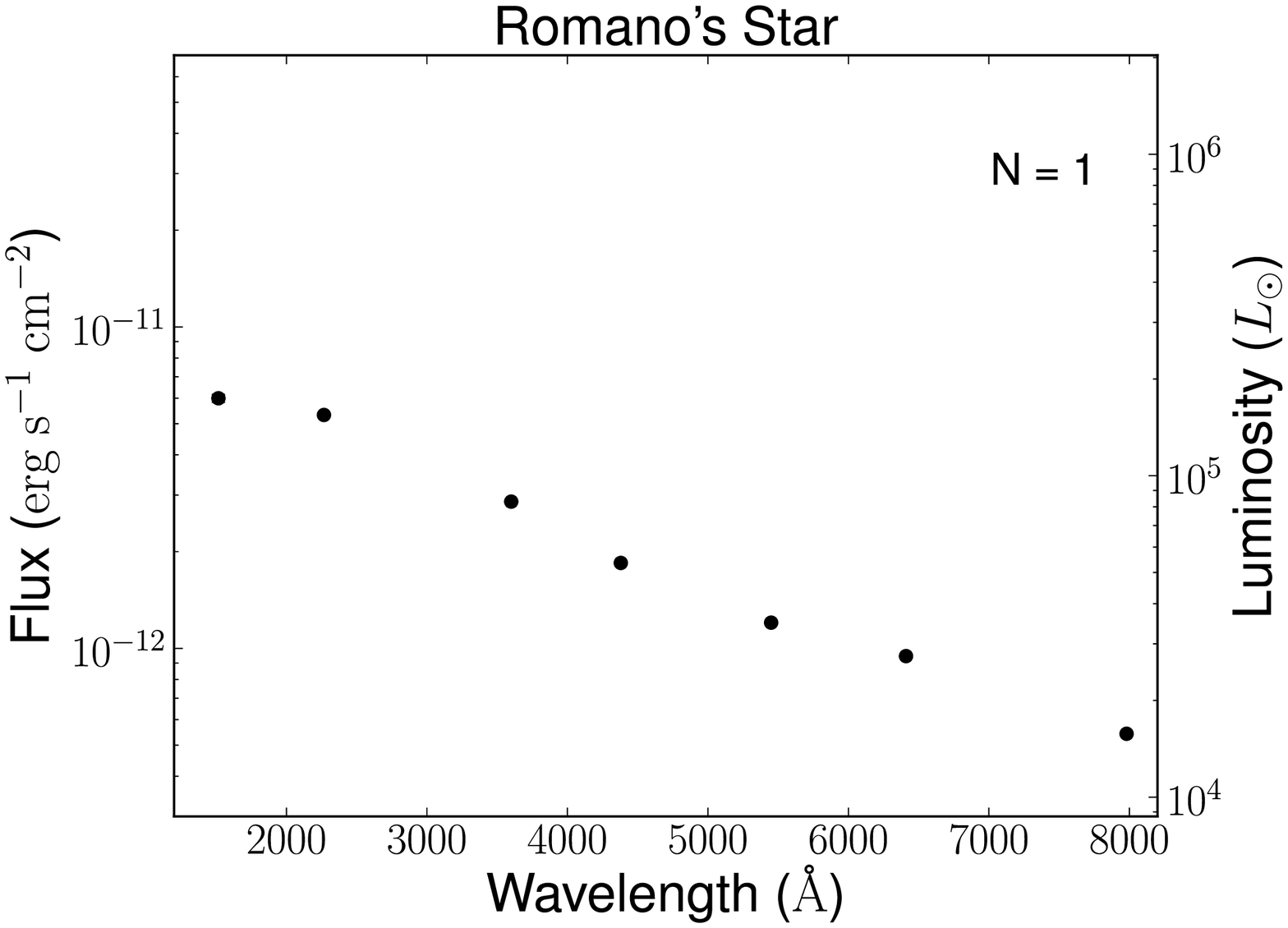}
  }
  \centerline{
    \includegraphics[width = 7.0cm]{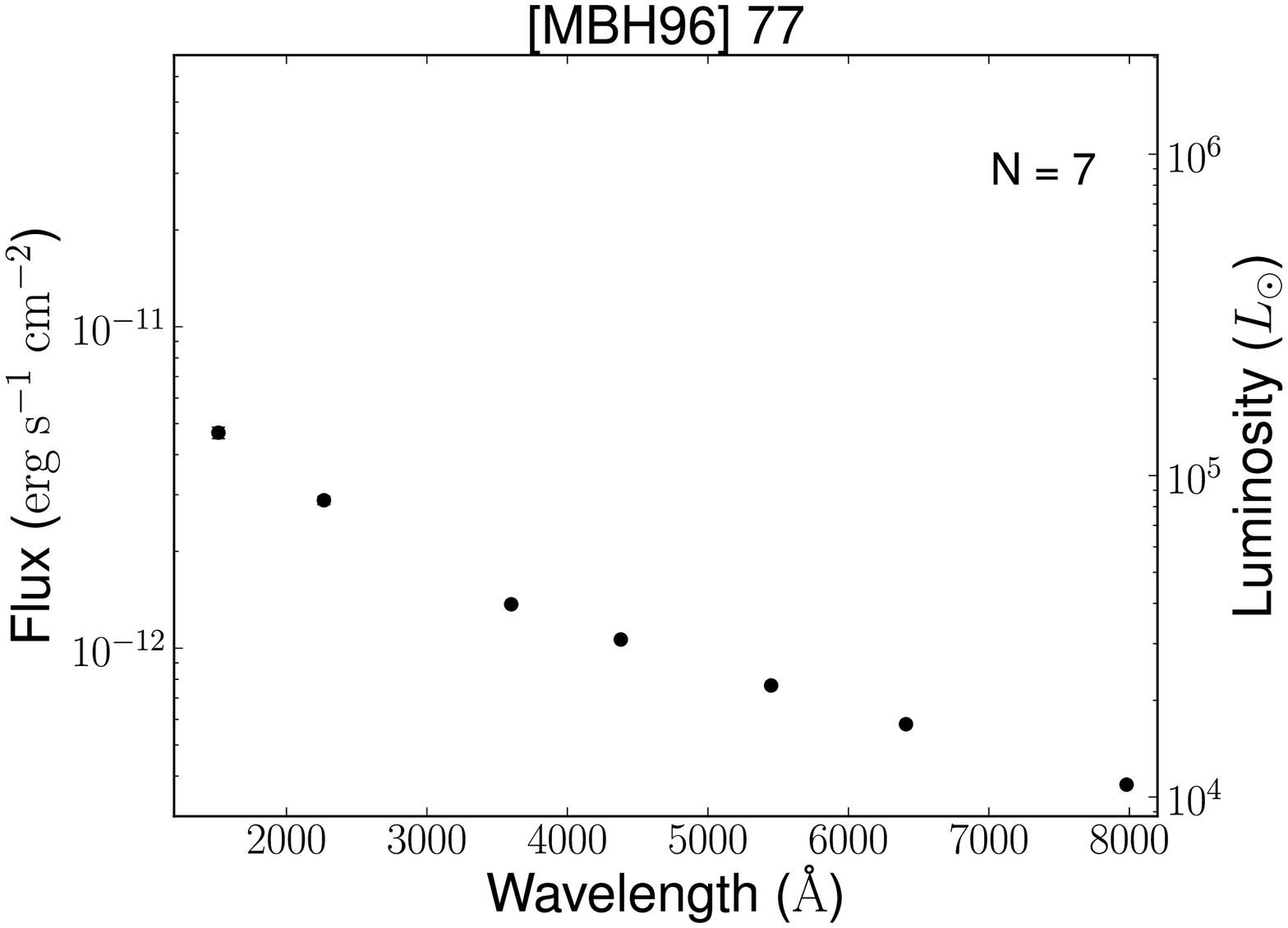}
    \includegraphics[width = 7.0cm]{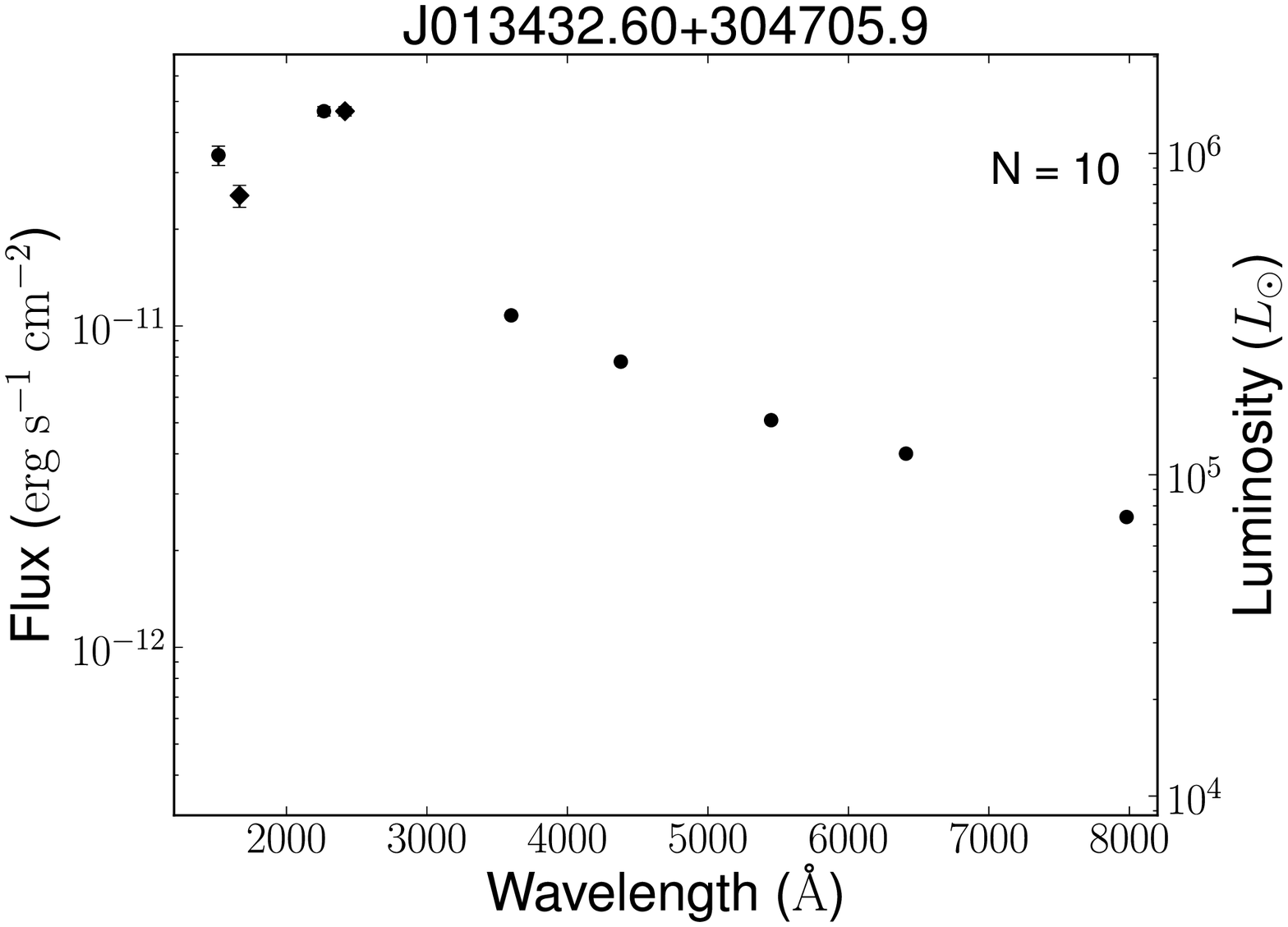}
  }
  \centerline{
    \includegraphics[width = 7.0cm]{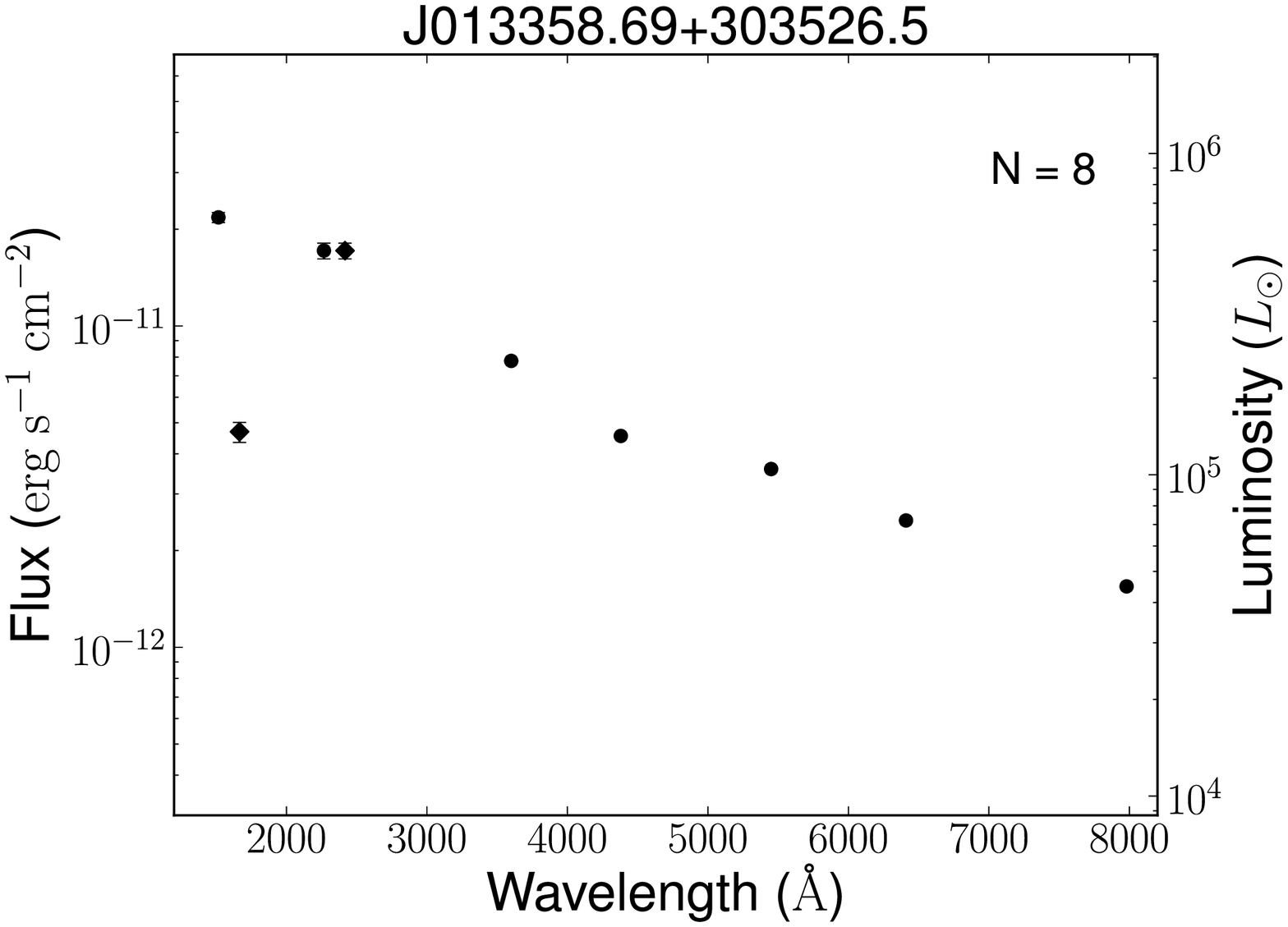}
    \includegraphics[width = 7.0cm]{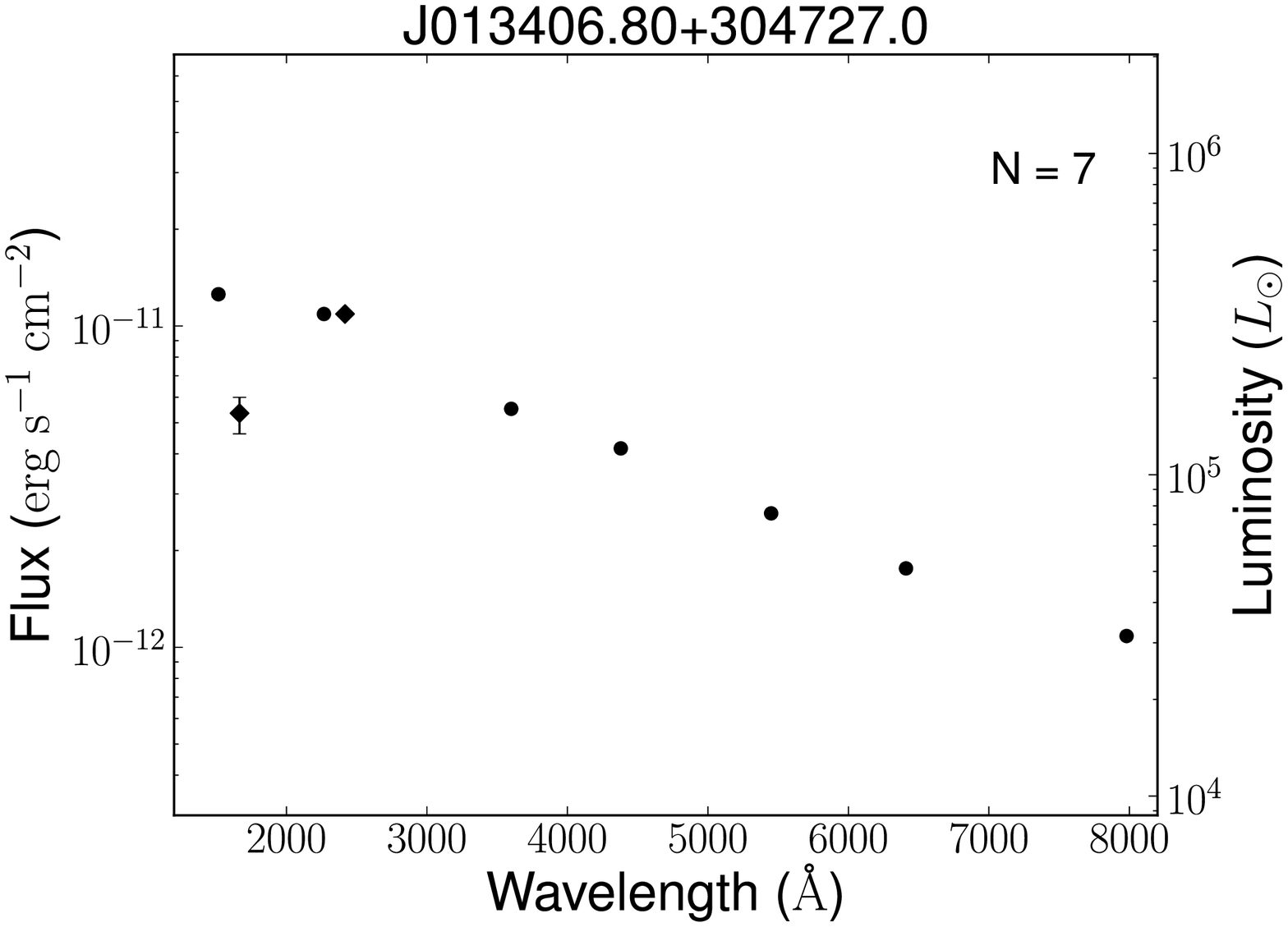}
  }
  \caption{A selection of some of the UV-brightest and most well-studied Wolf-Rayet stars in M33 that are in the final catalog.  \emph{First Row}:  [MBH96] 16 (left) and Romano's Star (right).  \emph{Second Row}:  [MBH96] 77 (left) and J013432.60+304705.9 (right).  \emph{Third Row}:  LGGS J013358.69+303526.5 (left) and J013406.80+304727.0 (right).  In the top-right corner of each panel, the value ``N'' corresponds to the number of optical sources that were within $3.^{\!\!\prime\prime}2$ of the UV source and is provided as a measure of crowding, which tended to be most severe for the few hundred brightest of the UV sources.  In the cases where there are multiple UV sources with a given optical star as its best match, the value given for ``N'' is the average amongst these sources.  For a description of the matching process, see \textsection\ref{subsec: opticalmethod}.}
  \label{fig: WR_SEDs}
  
\end{figure*}

With our UV/optical catalog in hand, we next investigate a selection of the more well-known stars in M33 that are likely also UV bright.  First, we look at a selection of Wolf-Rayet stars (WR), thought to be massive ($>$20$\msol$) stars that have blown off their envelopes and have become essentially hot, exposed stellar cores.  With effective temperatures typically in excess of 30,000 K, occasionally reaching well over 100,000 K, these stars should peak at a wavelength blueward of even GALEX's FUV filter.  As such, ignoring any absorption or extinction, we expect these stars to be among the highest luminosity in the FUV from the bandpasses presented here.  In Figure \ref{fig: WR_SEDs}, the sources [MBH96] 16, Romano's Star (\citealp{Romano78}), and [MBH96] 77, presented in the top and middle rows, are a selection of the most well-known WR stars in the galaxy.  [MBH96] 16, as can be seen from the figure, has three separate UV matches to its optical counterpart in our data.  Two of these peak in the NUV, whereas one continues to rise in the FUV.  Since it is a WR star, the most likely true match is this hottest UV source.  Romano's Star, [MBH96] 77, LGGS J013358.69+303526.5, and LGGS J013406.80+304727.0 also exhibit this behavior, albeit with somewhat different slopes.  Comparing our sources to a catalog of known WR stars in M33 \citep{Neugent11}, we recover \wrMatches\textrm{} out of a total 206.

\begin{figure*}
\centerline{
    \includegraphics[width = 7.0cm]{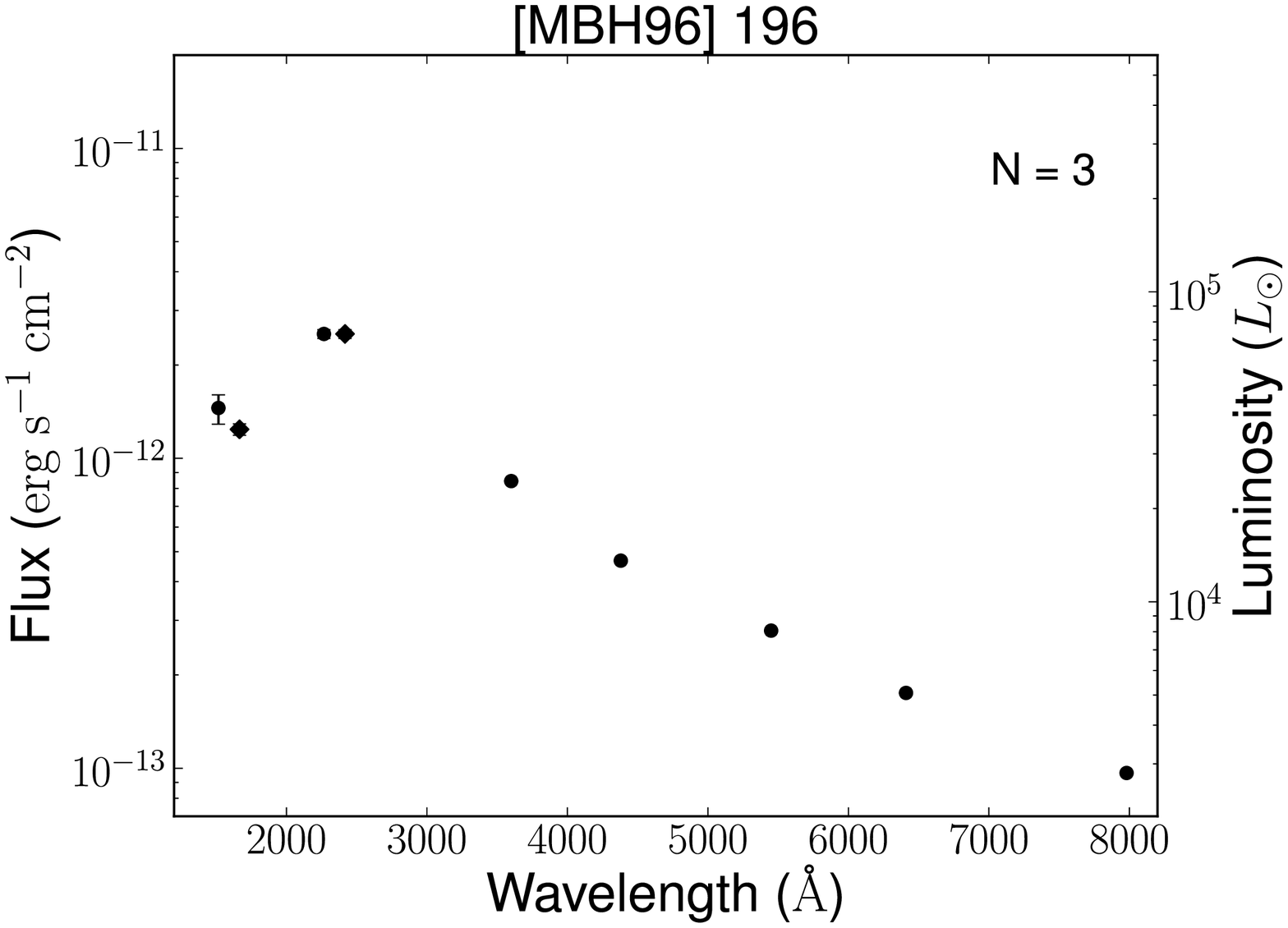}
    \includegraphics[width = 7.0cm]{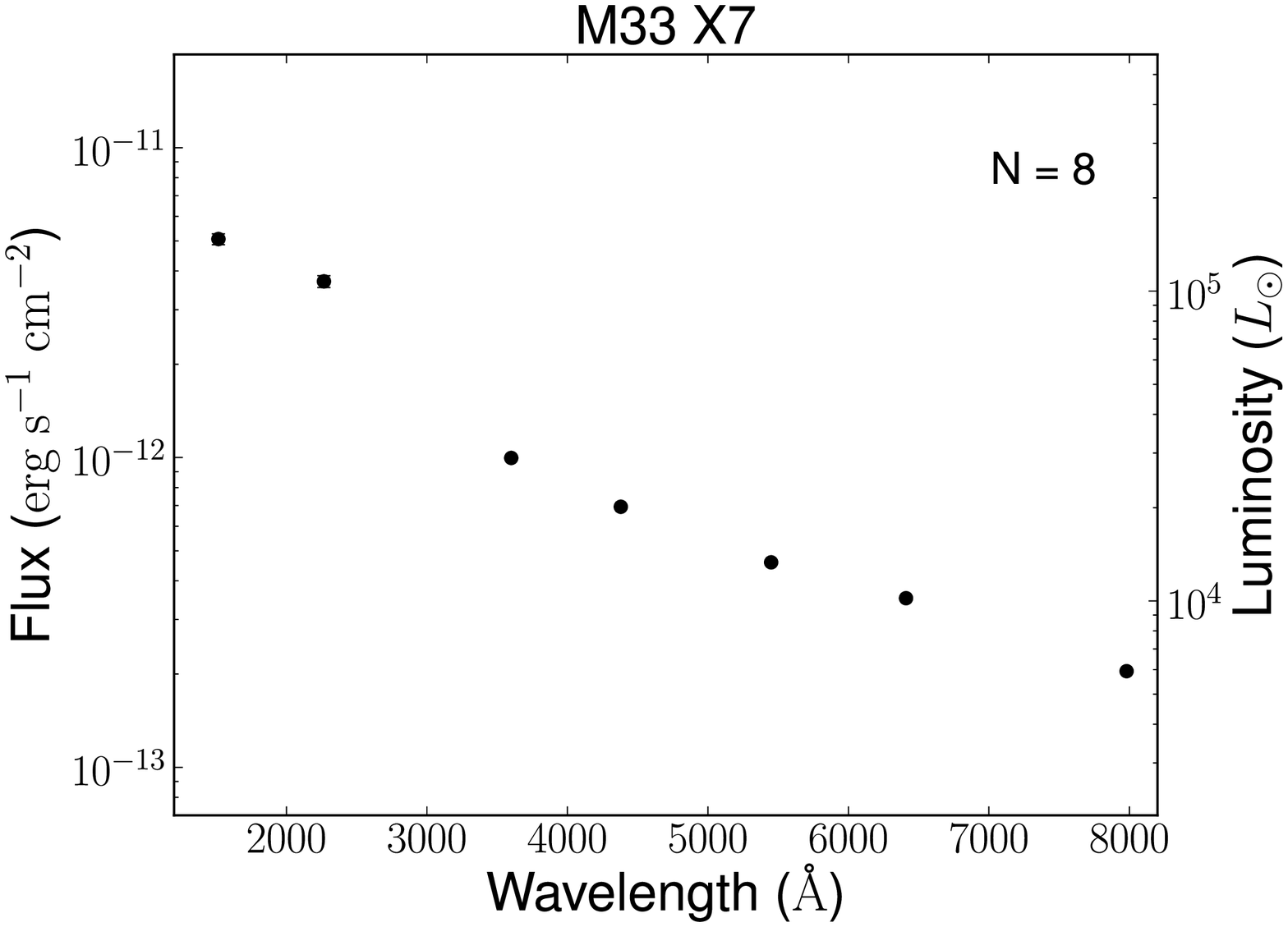}
   } 
  \centerline{
    \includegraphics[width = 7.0cm]{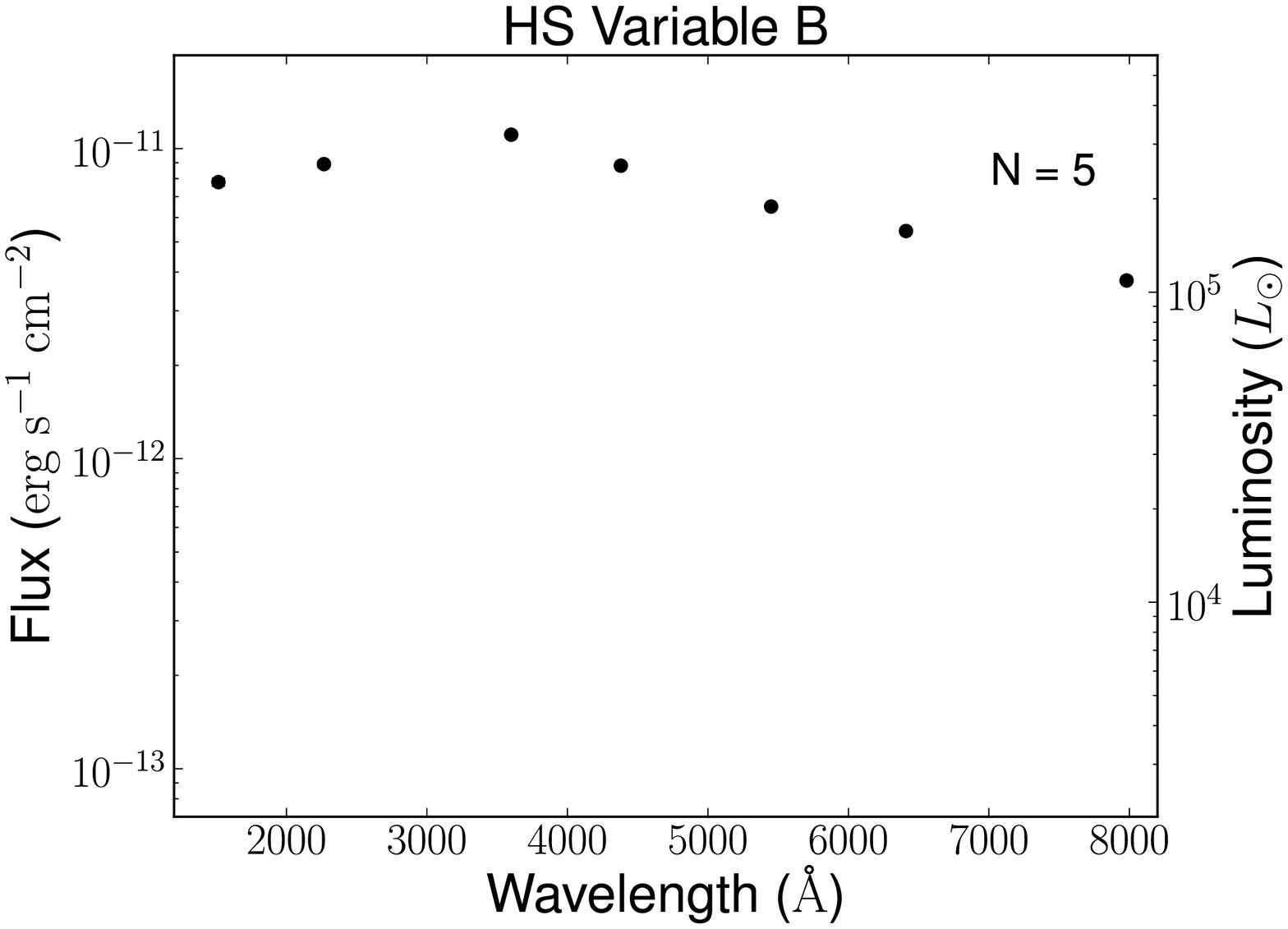}
    \includegraphics[width = 7.0cm]{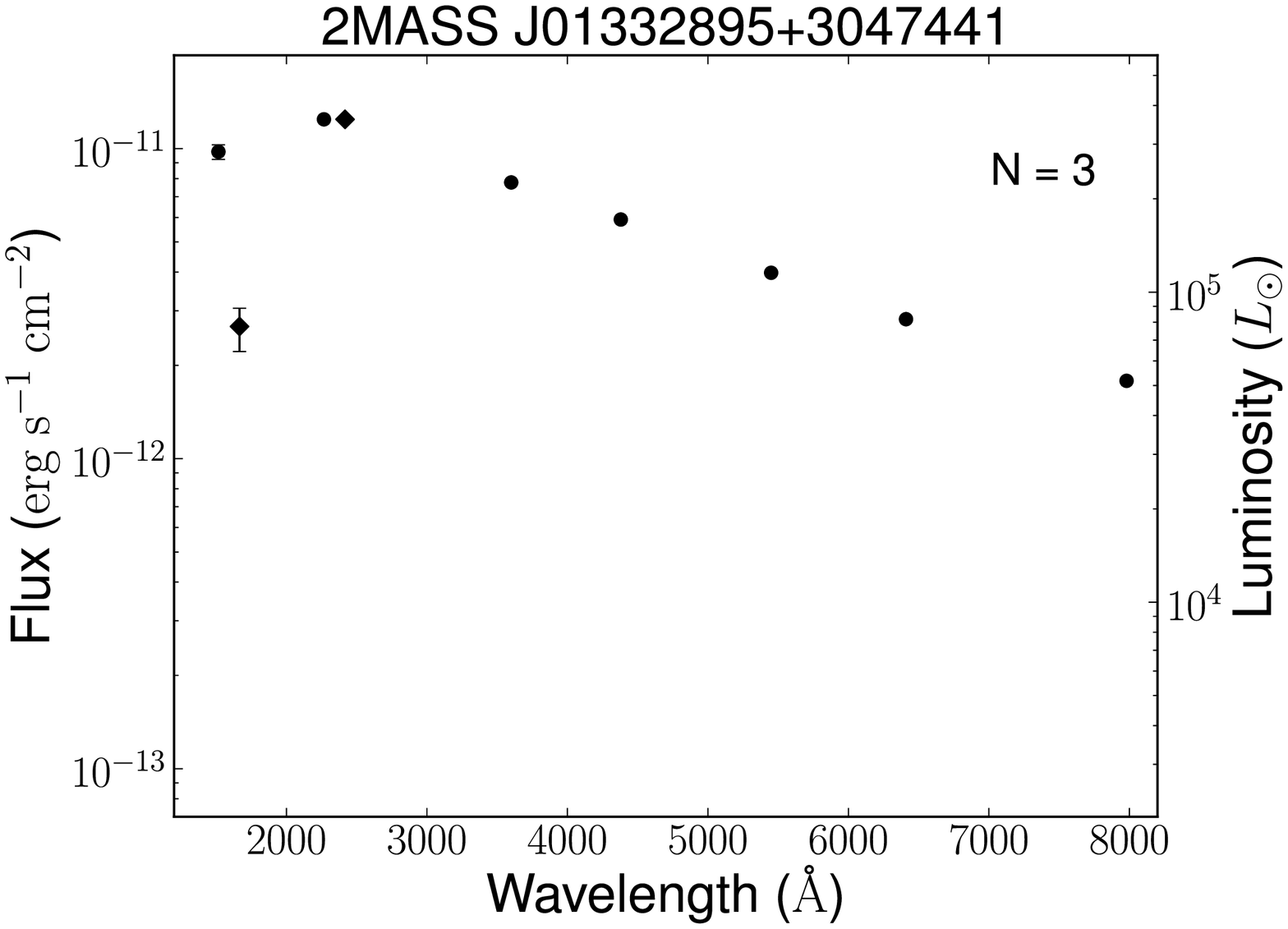}
  }
\caption{Same as Figure \ref{fig: WR_SEDs} but for other, non WR sources in M33.  \emph{First Row}:  [MBH96] 196, a detached eclipsing binary that has been used to measure the distance to M33, and M33 X-7 (HMXB).  \emph{Second Row}:  Hubble-Sandage Variable B (left) and 2MASS J01332895+3047441, both blue supergiants.}
  \label{fig: Other_Interesting_SEDs}
\end{figure*}

Another star we investigate is a detached eclipsing binary used by the DIRECT Project \citep{Bonanos06} to measure the distance to M33 itself, whose SED is presented in Figure \ref{fig: Other_Interesting_SEDs}.  The two stars in this binary have derived temperatures in excess of 35,000 K.  Our best matches, however, peak in the near-UV rather than the FUV, as one might expect.  Next, we look at M33 X-7 (e.g, \citealp{Long81, Long10}), a high mass X-ray binary.  From Figure \ref{fig: Other_Interesting_SEDs}, we see that it is quite luminous in the UV bands, as might be predicted due to the presence of both an accreting compact object and a hot stellar companion.    We also have a number of well-studied blue supergiants in our sample, including Hubble-Sandage Variable B \citep{Hubble53} and 2MASS J01332895+3047441 \citep{Kunchev86, Ivanov93, Skrutskie06}.  These stars, although blue, are much cooler than WR stars and we anticipate their spectral energy distributions to peak at longer wavelengths.  Both matches to 2MASS J01332895+3047441 peak in the NUV.  This star has been classified as a B-type supergiant \citep{Massey06}, which can indeed be warm enough to peak blueward of the optical range.  

\begin{figure*}
  \centerline{
    \includegraphics[width = 7.0cm]{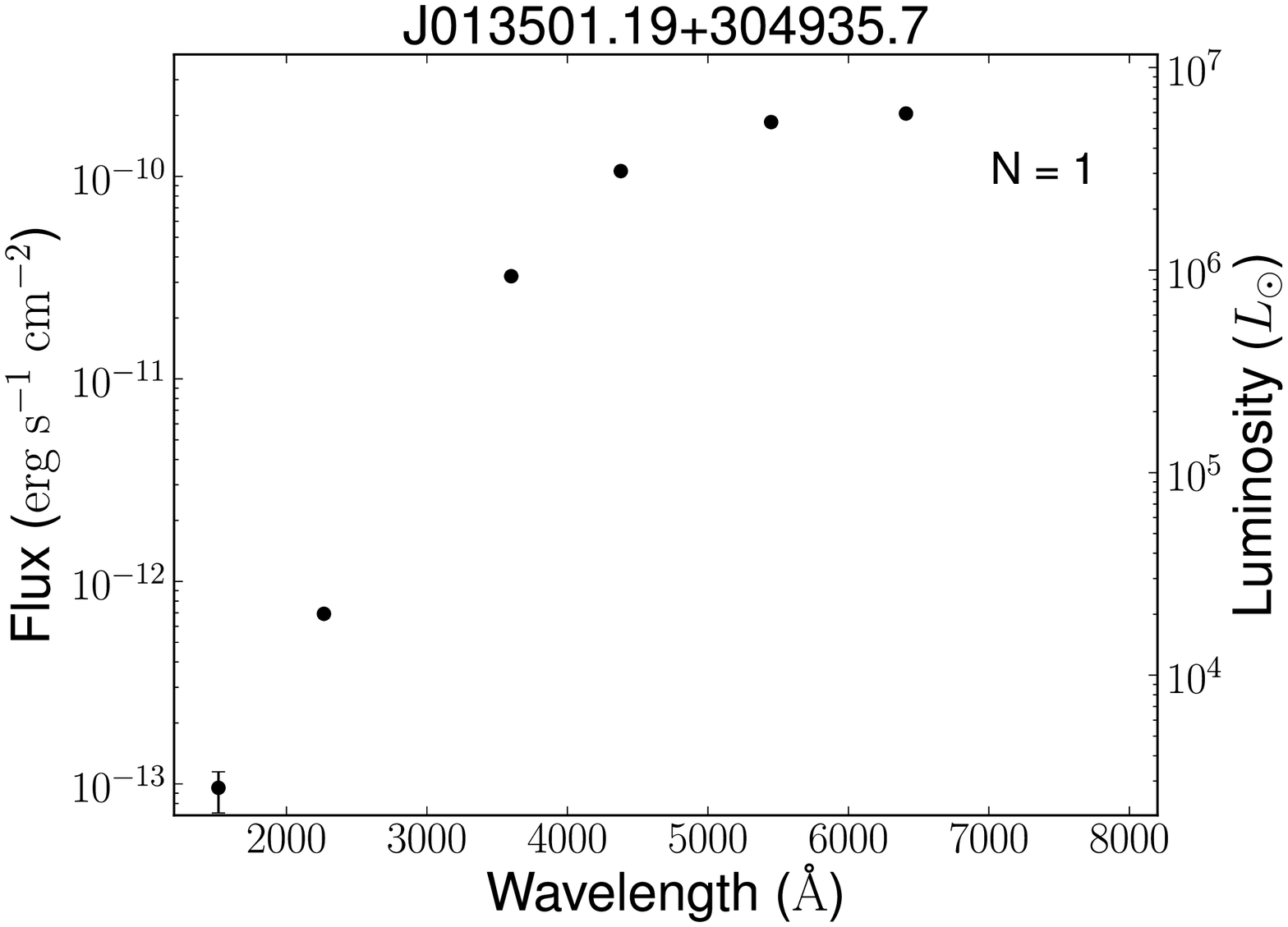}
    \includegraphics[width = 7.0cm]{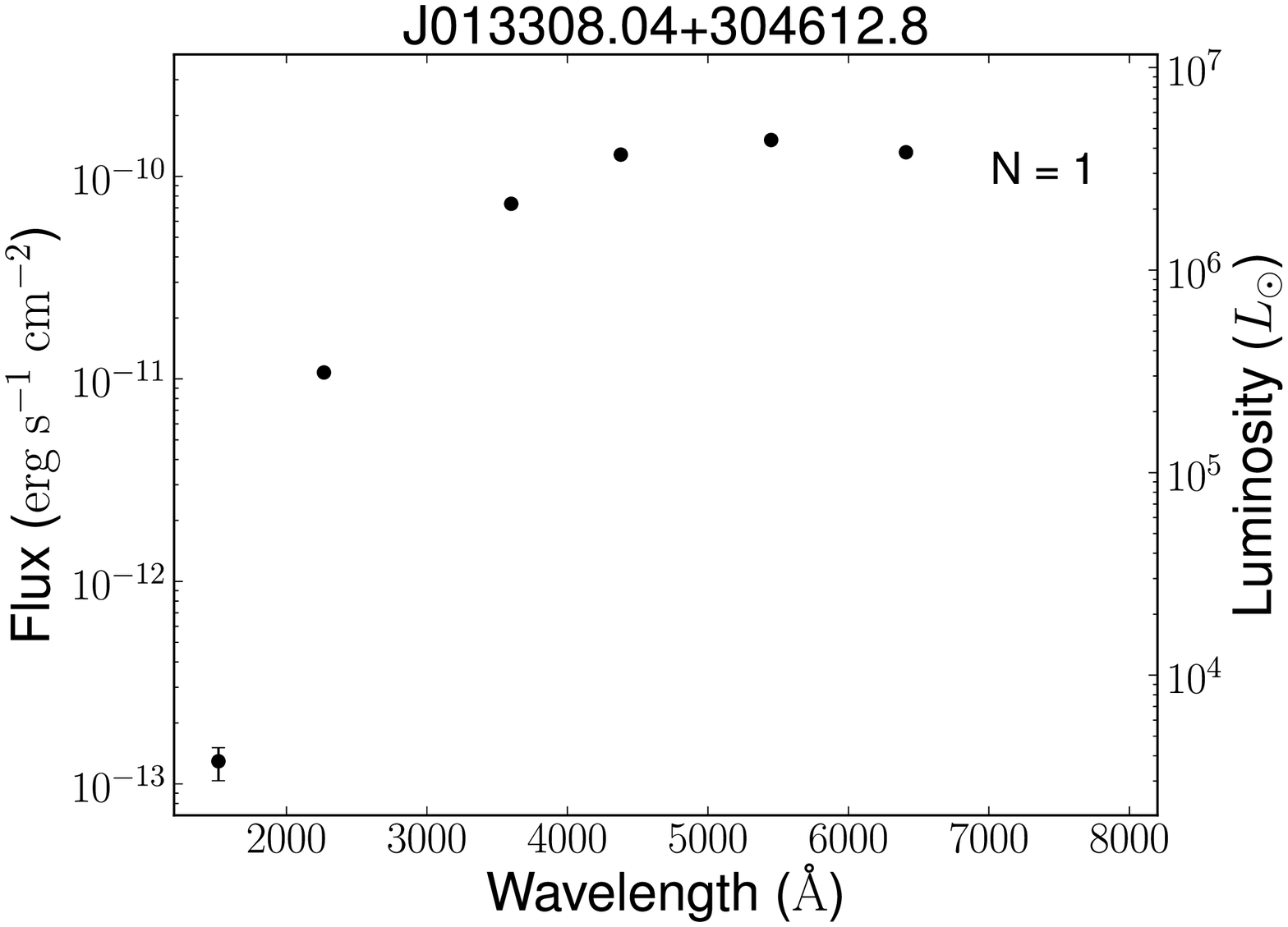}
  }
  \centerline{
    \includegraphics[width = 7.0cm]{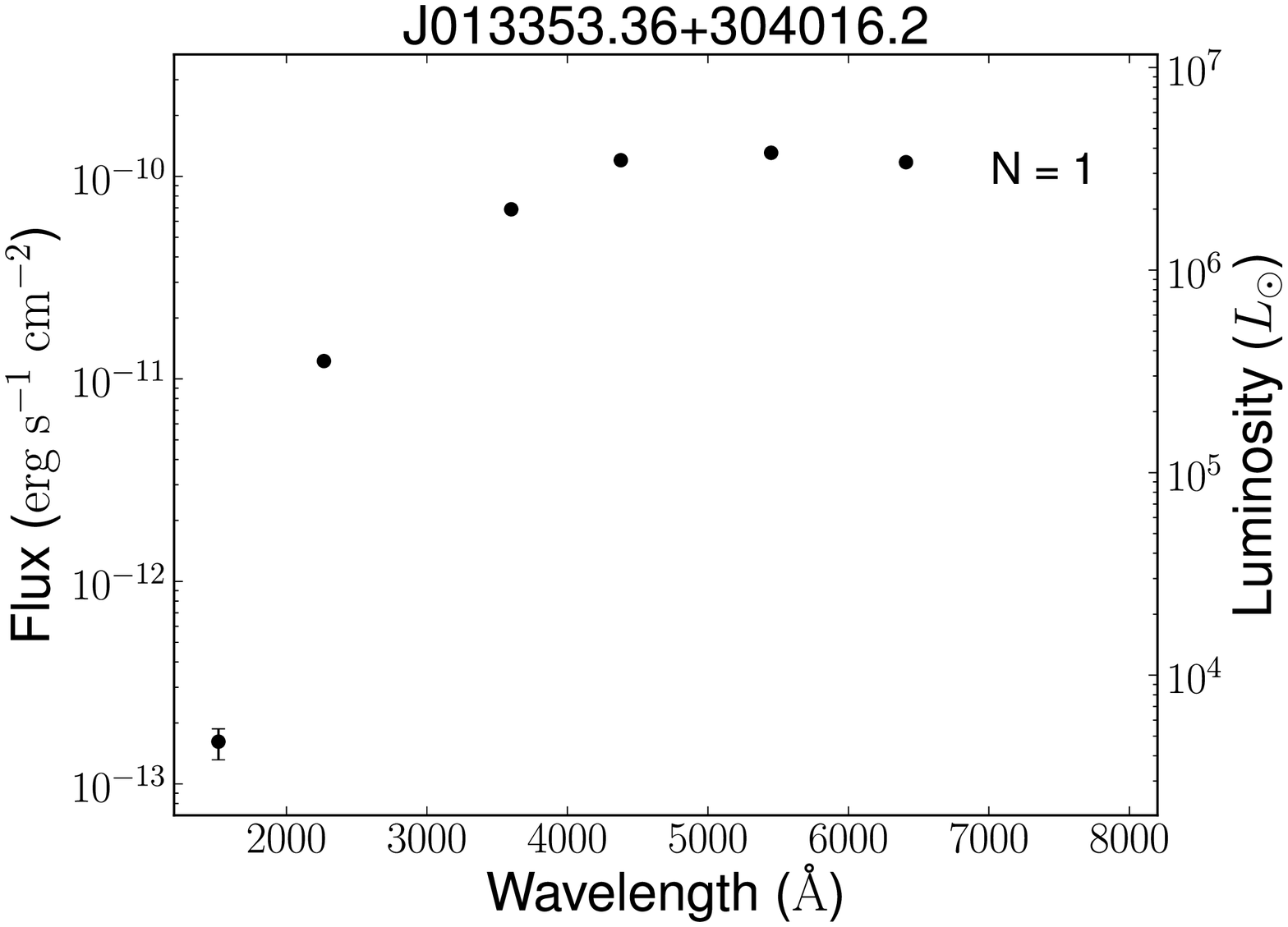}
    \includegraphics[width = 7.0cm]{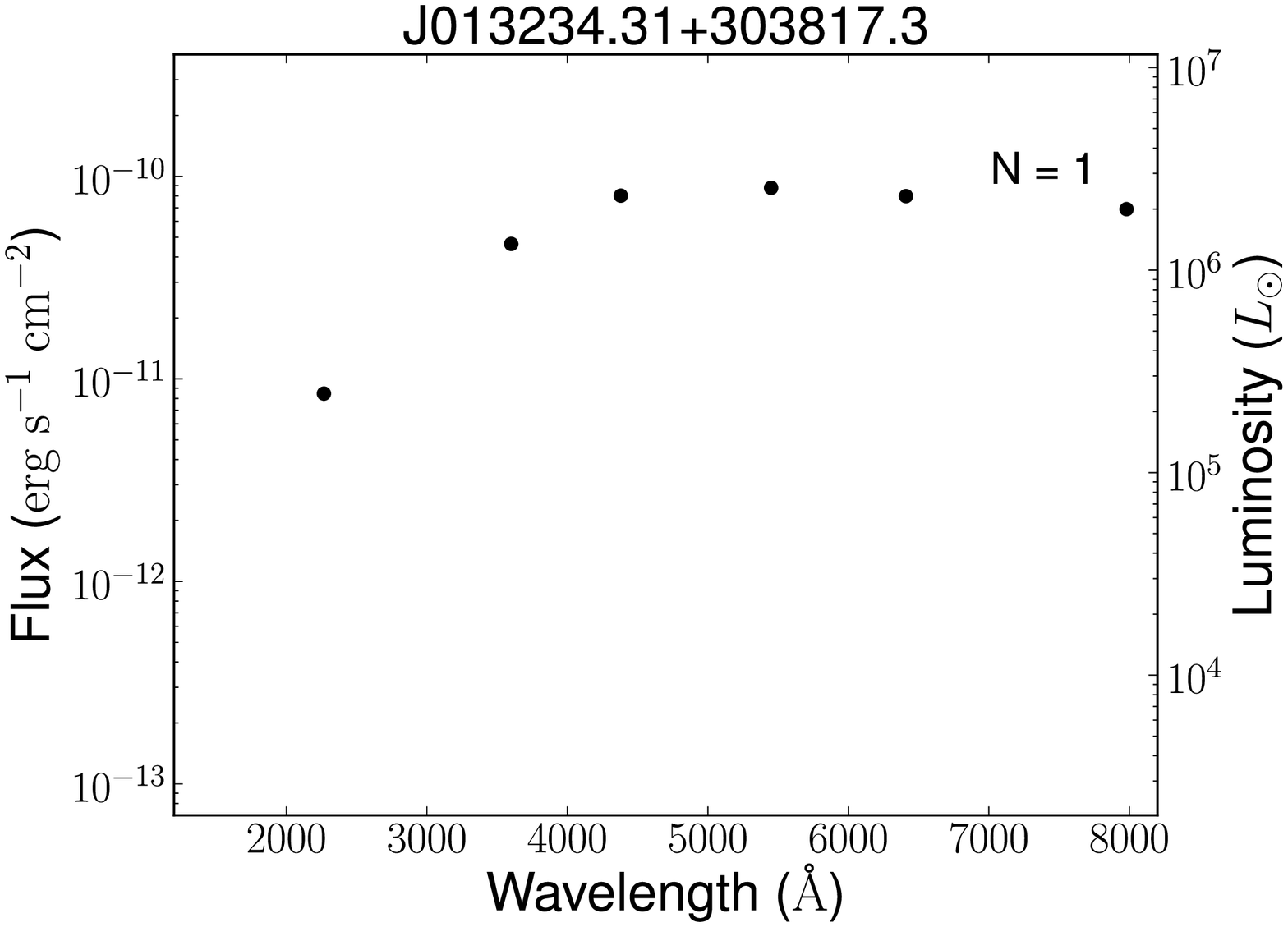}
  }
  \centerline{
    \includegraphics[width = 7.0cm]{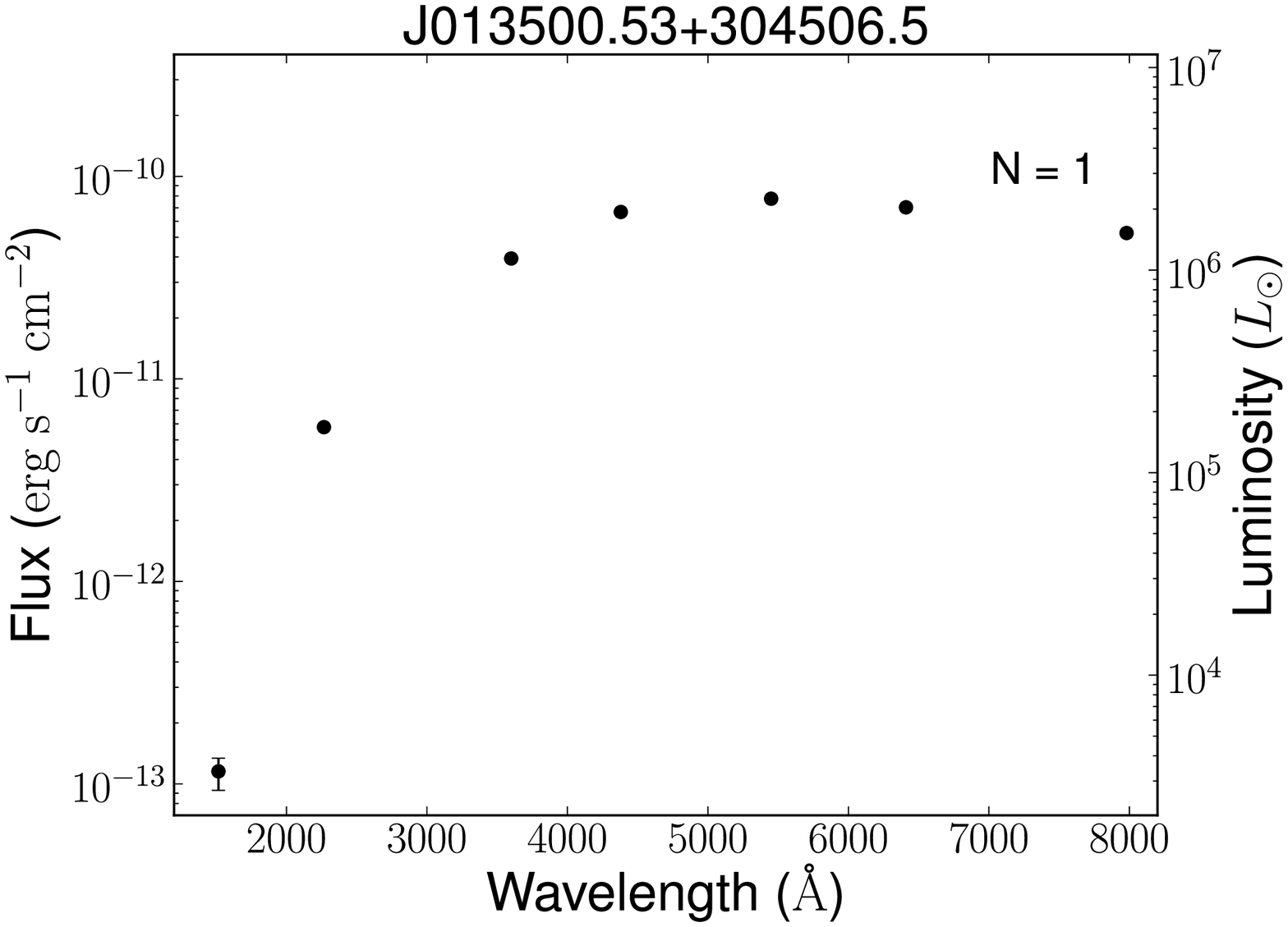}
    \includegraphics[width = 7.0cm]{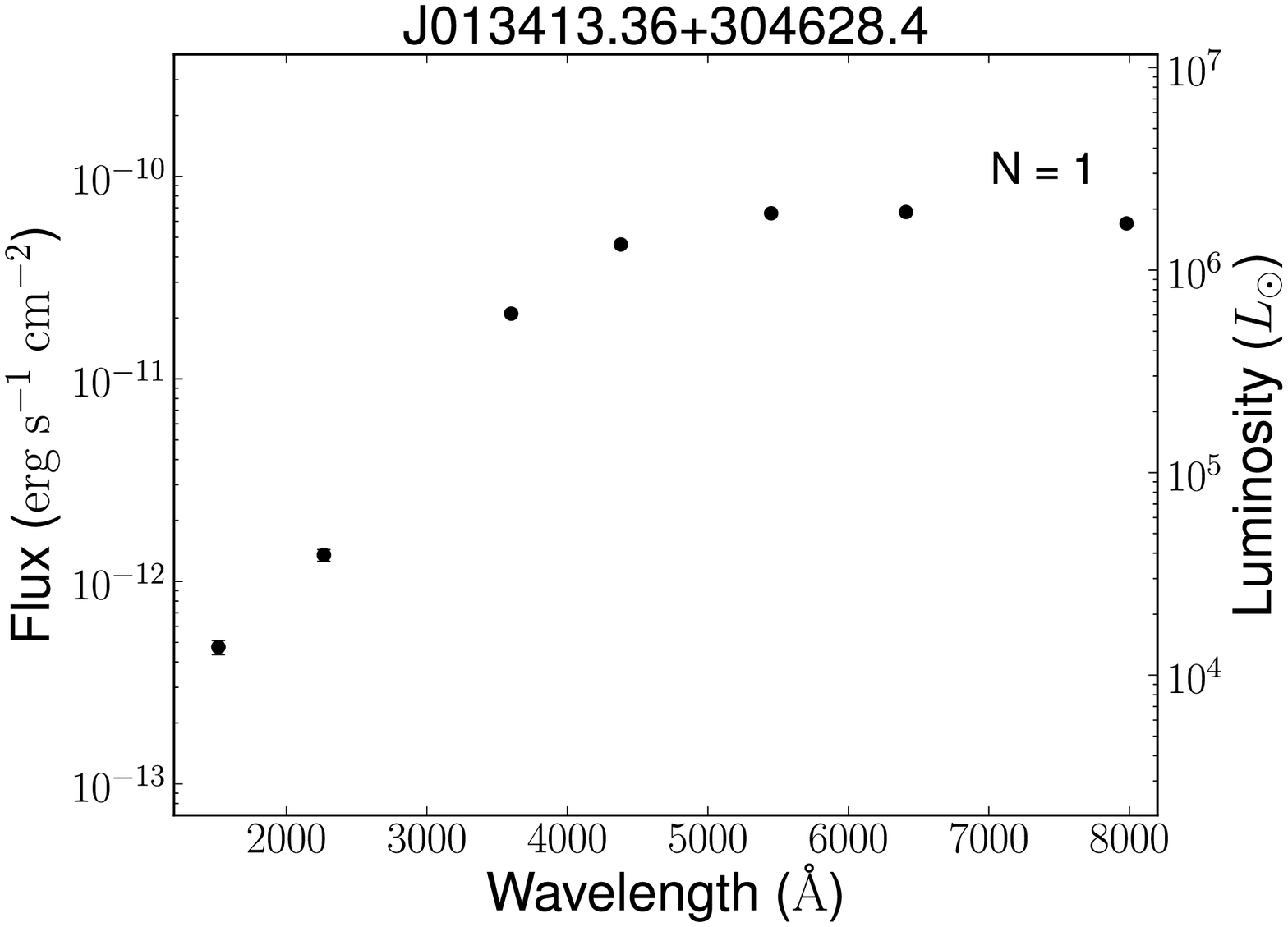}
  }
  \caption{Same as Figure \ref{fig: WR_SEDs} but for the six most luminous stars in the catalog (combination of all bands).}
  \label{fig: Brightest6_SEDS}
\end{figure*}

After searching the catalog for known interesting sources, an obvious next step is to look at the luminous stars in the catalog.  We present the 6 most luminous sources in Figure \ref{fig: Brightest6_SEDS}, where ``most luminous'' here is defined as the summation of fluxes from all available bands.  Immediately a trend appears.  All of these stars have most of their flux in the optical bandpasses and the UV data contributes minimally to their total bolometric luminosity.  The brightest stars in the catalog, then, are likely all relatively common and cooler red supergiants, though there may be also some of the rarer yet more intrinsically luminous yellow supergiants, as well as a few foreground sources.

\begin{figure*}
  \centerline{
    \includegraphics[width = 7.0cm]{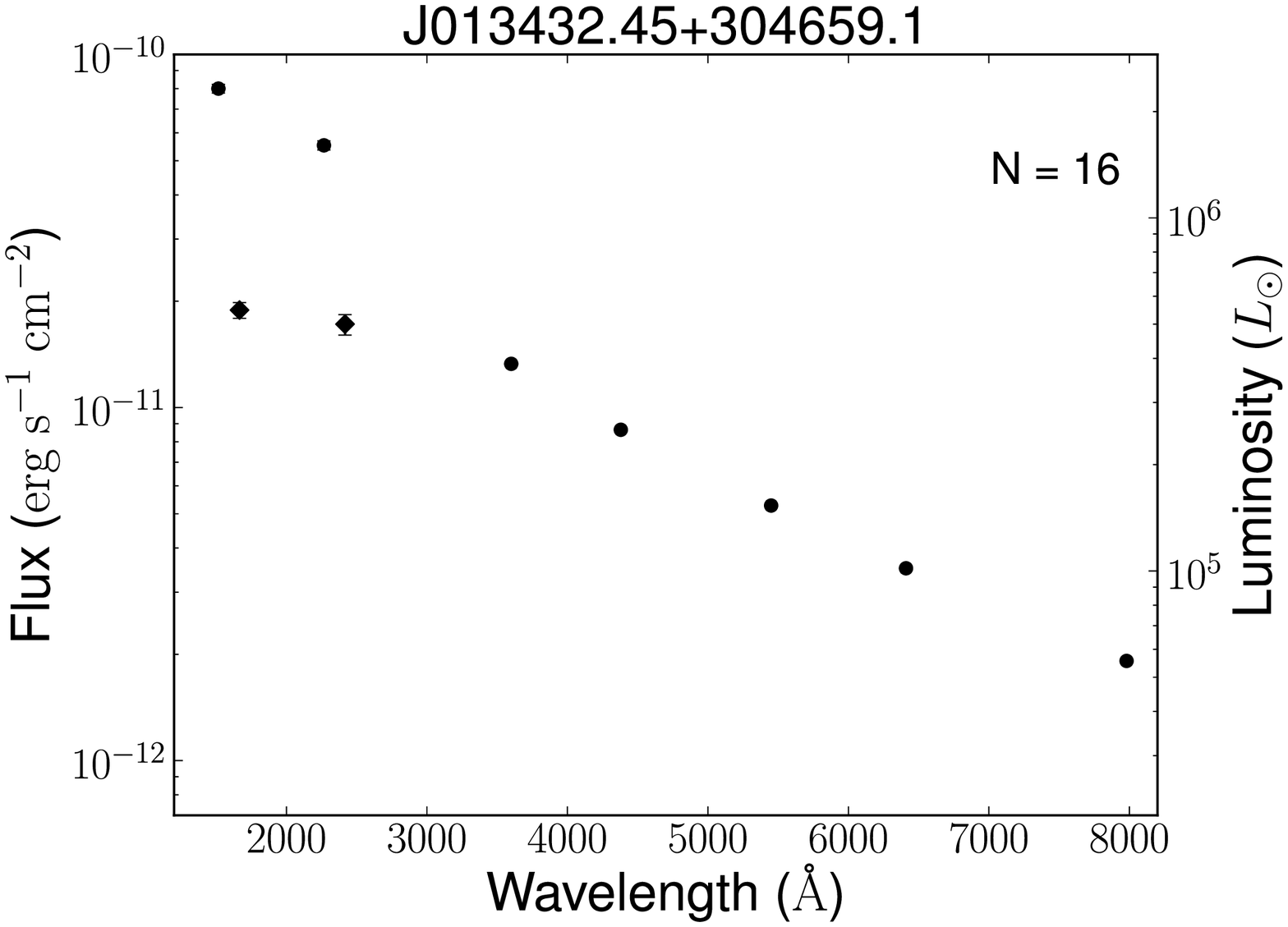}
    \includegraphics[width = 7.0cm]{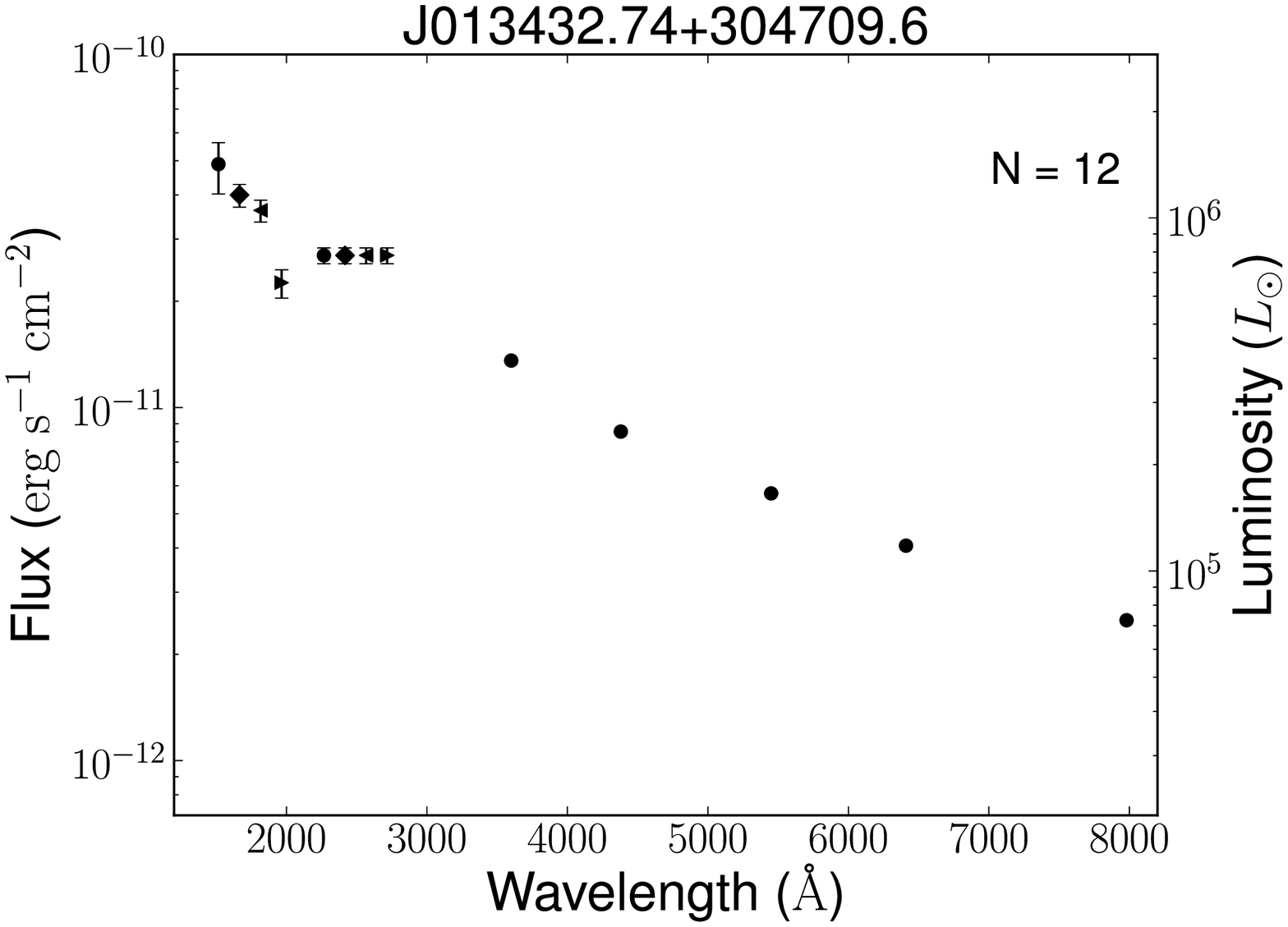}
  }                                 
  \centerline{    
    \includegraphics[width = 7.0cm]{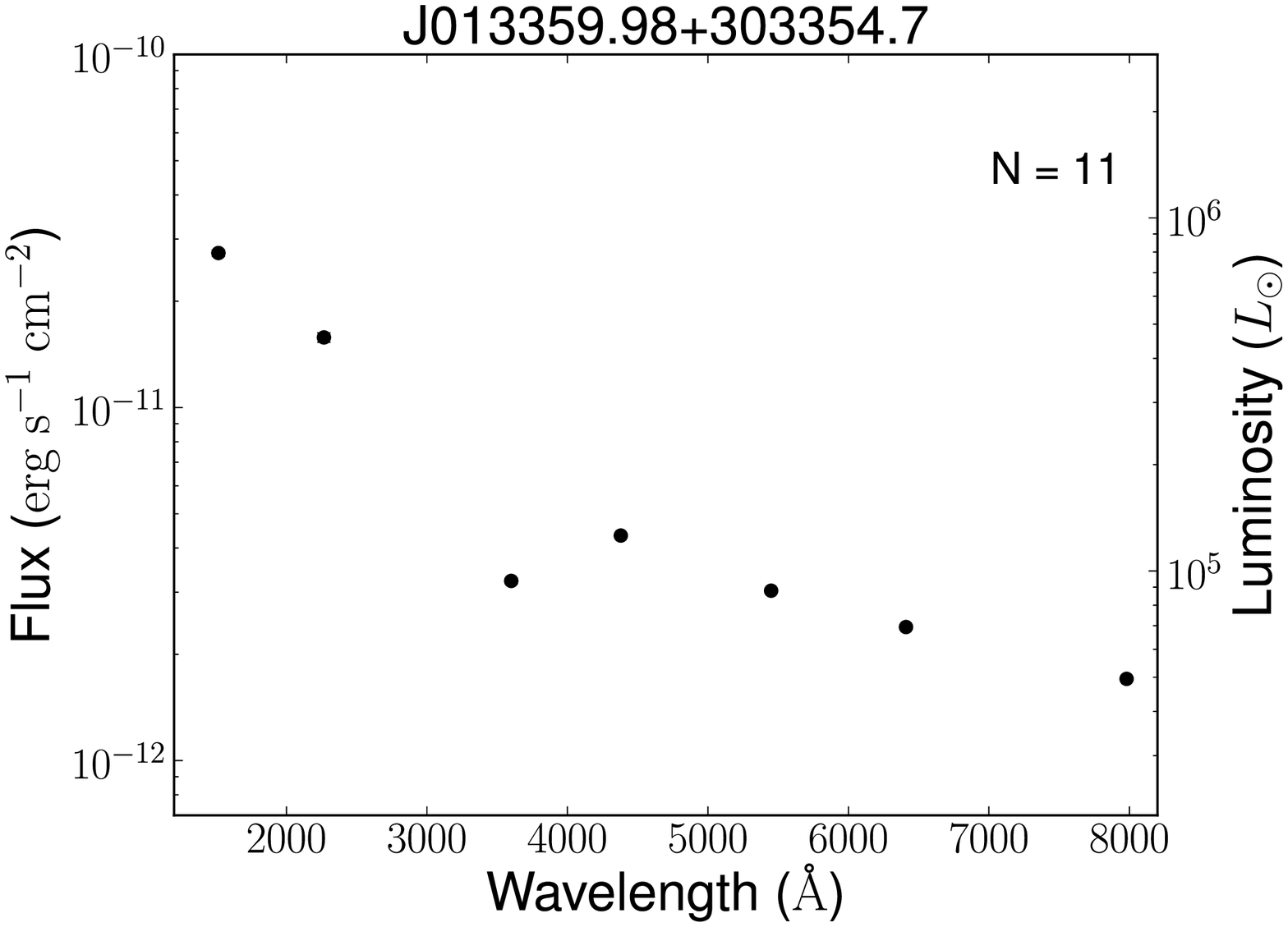}
    \includegraphics[width = 7.0cm]{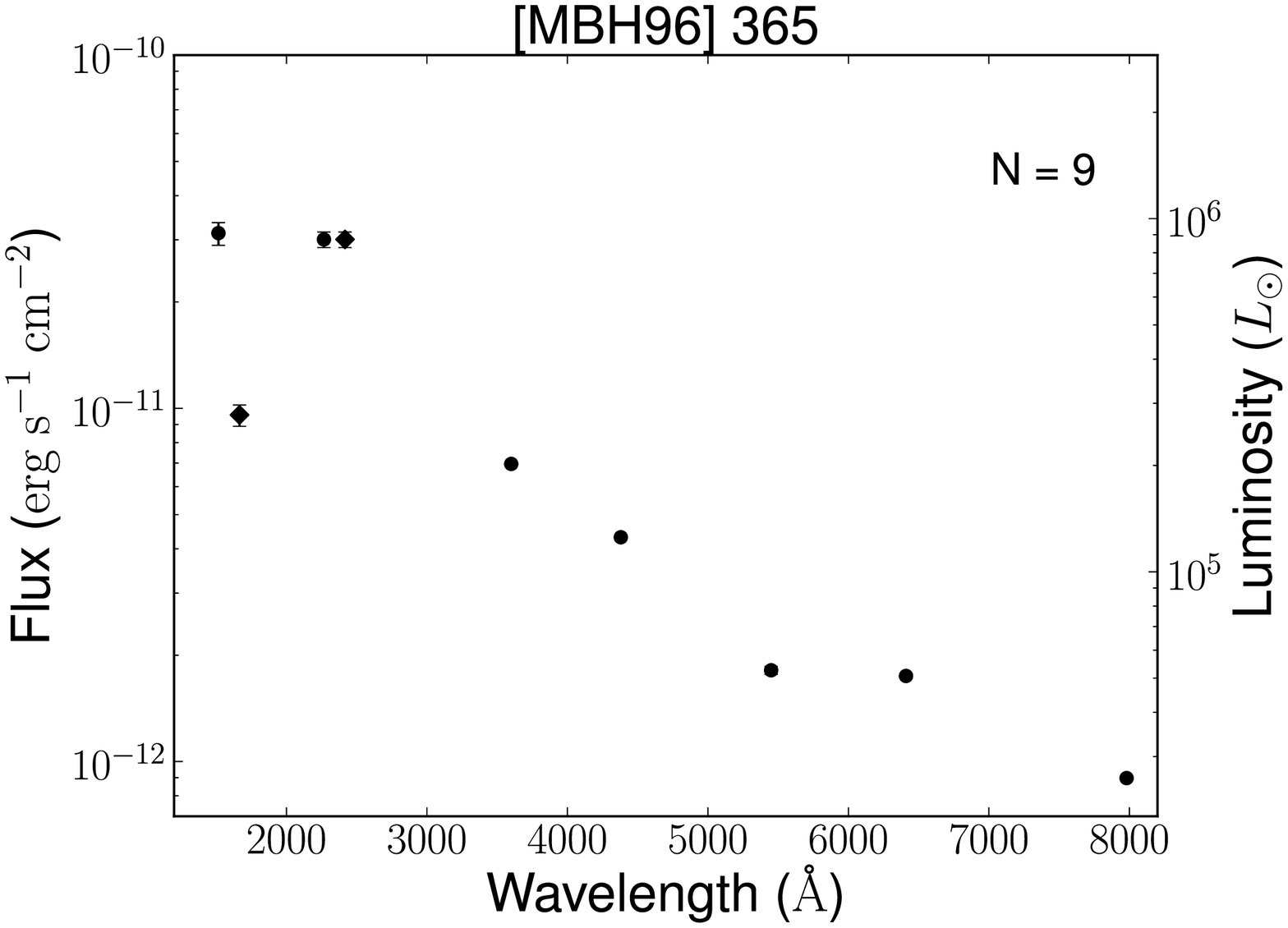}
  }
  \centerline{
    \includegraphics[width = 7.0cm]{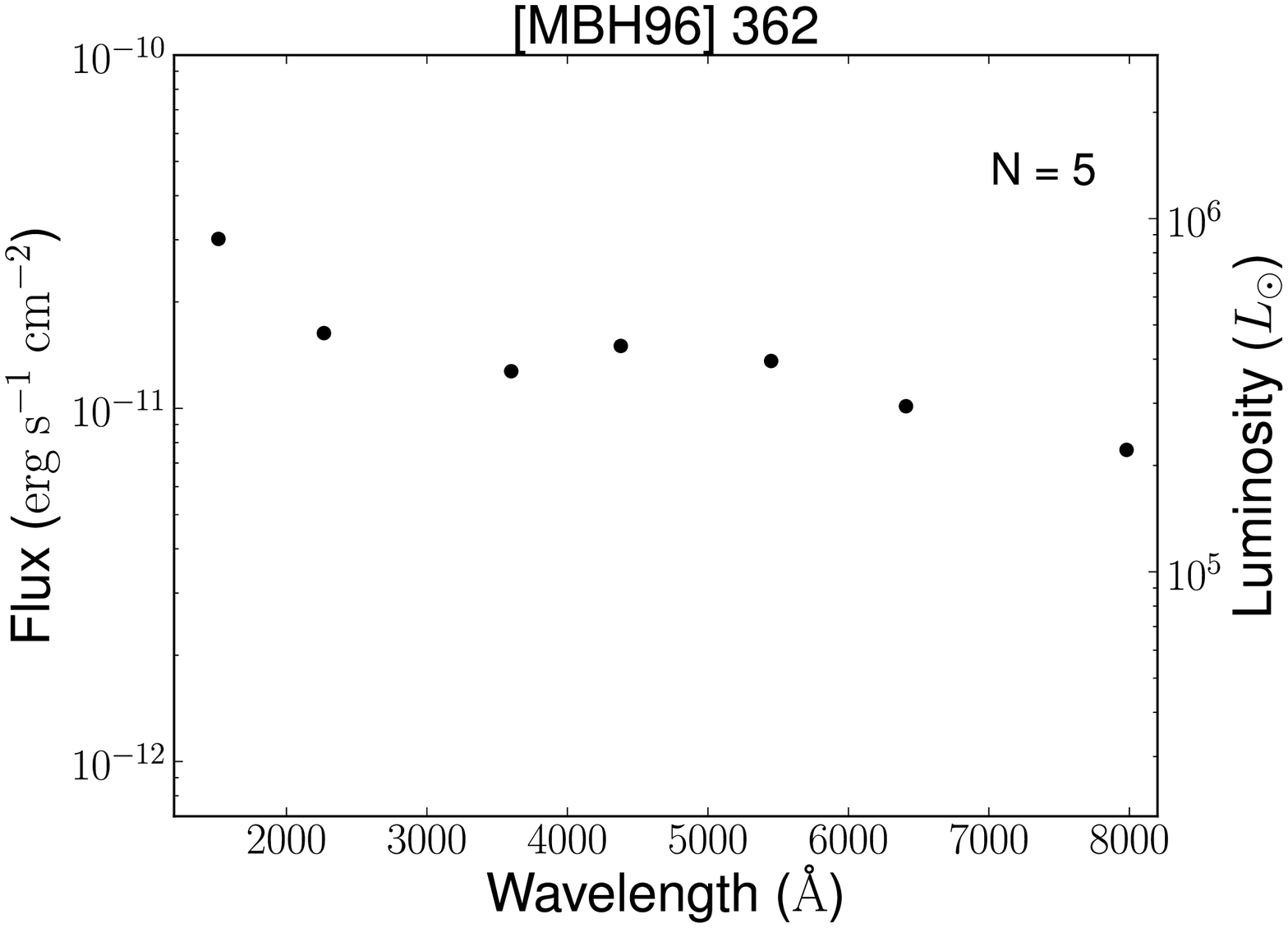}
    \includegraphics[width = 7.0cm]{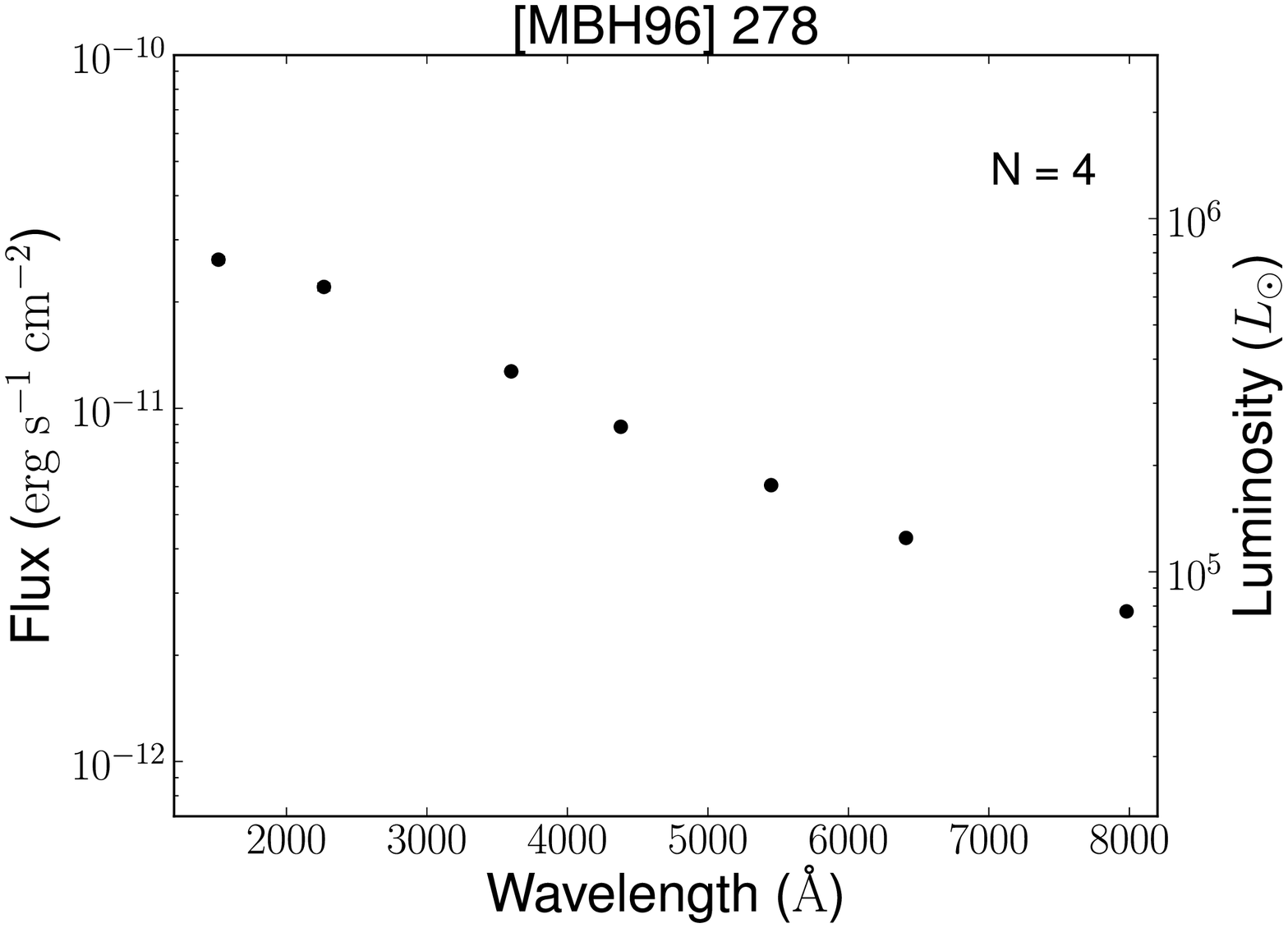}
  }
  \caption{Same as Figure \ref{fig: WR_SEDs} but for the six stars with the highest UV luminosity in the catalog.  We note that the UV brightest stars also tend to be the most crowded, as is evident from the large values of ``N''.}
  \label{fig: Brightest6UV_SEDS}
\end{figure*}

For which stars does the UV data significantly change their inferred bolometric luminosities?  To answer this question, we present the six stars with the highest luminosity in GALEX's FUV bandpass in Figure \ref{fig: Brightest6UV_SEDS}.  Three of these were found in the UIT catalog of M33, but three are new UV sources.  All of these tend to have near and far-UV luminosities of around $10^{6}\lsol$.  This is a factor of six down from the peak bandpass luminosity in Figure \ref{fig: Brightest6_SEDS}, but it is of the same order of magnitude as in those stars.  The difference is that the UV bright stars are bright solely in the UV, often dropping by at least a factor of two in flux between the near-UV and V bands.  And this is not unexpected of extremely hot stars which will emit significantly in the UV but have their flux fall continually in redder bands.

With this work, we seek to update and expand the existing UV catalog of M33 using archival GALEX data.  Using PSF photometry to optimally find and photometer sources, we find tens of thousands of more sources than in the pipeline product as it was not constructed to handle such a crowded environment.  We match these to the UIT catalog \citep{Massey96} and recover all but eight of these sources.  Next, we match to an optical catalog from the LGGS (\citealp{Massey06}) to create a final catalog with \numOurSources sources, of which \numoptuvmatches\textrm{} have optical matches thereby spanning seven filters from the far-UV to near-IR.  We then investigate the properties of our catalog and find that the most overall luminous sources are typically brightest in the optical bands (and hence likely evolved stars), but there are still many sources, likely young, massive stars and WR stars, that are continuing to rise in the UV range, indicating high effective temperatures.

A useful future endeavor would be to perform a similar analysis with the \emph{Swift} UV data of M33 \citep{Immler08}.  \emph{Swift} covers 3 UV filters with high spatial resolution, which would further assist in lessening the obvious crowding issue that is persistent in both the GALEX and UIT data.  

\section*{Acknowledgements}
DM would like to thank Philip Massey, Scott Adams, Michael Fausnaugh, Rubab Khan, Ben Shappee, and Obright Lorain for helpful discussions that contributed greatly to this paper.  We also wish to thank the anonymous referee, who provided numerous useful comments and aided the betterment of this manuscript.

Based on observations made with the NASA Galaxy Evolution Explorer.  Some of the data presented in this paper were obtained from the Mikulski Archive for Space Telescopes (MAST). STScI is operated by the Association of Universities for Research in Astronomy, Inc., under NASA contract NAS5-26555. Support for MAST for non-HST data is provided by the NASA Office of Space Science via grant NNX13AC07G and by other grants and contracts.

\bibliographystyle{mn2e}
\bibliography{UVreferences}

\end{document}